\documentclass[draft]{agujournal2019}
\usepackage{url} 
\usepackage{lineno}
\usepackage[footnotes]{trackchanges} 
\usepackage{soul}

\usepackage{comment}
\usepackage{multirow}
\usepackage{amsmath}
\usepackage{float}
\usepackage{graphicx}
\usepackage{bm} 
\usepackage{mfirstuc}
\usepackage{upgreek} 
\usepackage{lscape} 
\usepackage{xcolor} 
\usepackage{ragged2e} 
\usepackage[capitalise]{cleveref}
\usepackage[export]{adjustbox} 

\newcommand{\filtered}[1]{\overline{{#1}}}
\newcommand{\GMcoeff}[2]{\ensuremath{\frac{\Delta^2}{#1\pm#2}}}

\newcommand{\apriori}{\textit{a priori }}
\newcommand{\aposteriori}{\textit{a posteriori }}

\newcommand{\KJ}[1]{{\color{black}#1}}
\newcommand{\KJnew}[1]{{\color{black}#1}}

\usepackage[acronym]{glossaries}
\newacronym{fhit}{FHIT}{forced homogeneous isotropic turbulence}
\newacronym{hit}{HIT}{homogeneous isotropic turbulence}
\newacronym{les}{LES}{large-eddy simulation}
\newacronym{dns}{DNS}{direct numerical simulation}
\newacronym{fdns}{FDNS}{filtered direct numerical simulation}
\newacronym{1d}{1D}{one-dimensional}
\newacronym{2d}{2D}{two-dimensional}
\newacronym{3d}{3D}{three-dimensional}
\newacronym{sgs}{SGS}{subgrid-scale}
\newacronym{rbc}{RBC}{Rayleigh-B\'enard convection}
\newacronym{rvm}{RVM}{relevance vector machine}
\newacronym{ml}{ML}{machine learning}
\newacronym{gm}{GM}{gradient model}
\newacronym{mse}{MSE}{mean-squared error}
\newacronym{gep}{GEP}{gene-expression programming}
\newacronym{cc}{CC}{correlation coefficients}

\everymath{\displaystyle} 

\justifying 

\DeclareUnicodeCharacter{2212}{-} 

\draftfalse

\journalname{Journal of Advances in Modeling Earth Systems (JAMES)}

\begin{document}


\title{Learning Closed-form Equations for Subgrid-scale Closures from High-fidelity Data: Promises and Challenges}

%
%

\authors{Karan Jakhar\affil{1,2}, Yifei Guan\affil{1,2}, Rambod Mojgani\affil{1,2},  Ashesh Chattopadhyay\affil{1,3}, and Pedram Hassanzadeh\affil{1,2}}

\affiliation{1}{Rice University, Houston, TX, USA}
\affiliation{2}{University of Chicago, Chicago, IL, USA}
\affiliation{3}{University of California, Santa Cruz, CA, USA}

\correspondingauthor{Pedram Hassanzadeh}{pedramh@uchicago.edu}

\begin{keypoints}
\item Subgrid-scale momentum/heat flux closures discovered using common algorithms are the analytically derivable nonlinear gradient model (NGM)
\item In 2D turbulence/convection, NGM leads to unstable online simulations due to its inability to fully capture key inter-scale energy transfers 
\item We suggest that physics-informed loss functions, libraries, \KJ{metrics, and sparsity selections} are needed to discover accurate/stable closures
\end{keypoints}

\glsresetall

\begin{abstract}
There is growing interest in discovering interpretable, closed-form equations for subgrid-scale (SGS) closures/parameterizations of complex processes in Earth systems. Here, we apply a common equation-discovery technique with expansive libraries to learn closures from filtered direct numerical simulations of 2D turbulence and Rayleigh-B\'enard convection (RBC). Across common filters \KJ{(e.g., Gaussian, box)}, we robustly discover closures of the same form for momentum and heat fluxes. These closures depend on nonlinear combinations of gradients of filtered variables, with constants that are independent of the fluid/flow properties and only depend on filter type/size. We show that these closures are the nonlinear gradient model (NGM), which is derivable analytically using Taylor-series. Indeed, we suggest that with common (physics-free) equation-discovery algorithms, \KJ{for many common systems/physics, discovered closures are consistent with the leading term of the Taylor-series (except when cutoff filters are used).} Like previous studies, we find that large-eddy simulations with NGM closures are unstable, despite significant similarities between the true and NGM-predicted fluxes (correlations $> 0.95$). We identify two shortcomings as reasons for these instabilities: in 2D, NGM produces zero kinetic energy transfer between resolved and subgrid scales, lacking both diffusion and backscattering. In RBC, potential energy backscattering is poorly predicted. Moreover, we show that SGS fluxes diagnosed from data, presumed the ``truth'' for discovery, depend on filtering procedures and are not unique. Accordingly, to learn accurate, stable closures in future work, we propose several ideas around using physics-informed libraries, loss functions, and metrics. These findings are relevant to closure modeling of any multi-scale system.

\end{abstract}

\glsresetall

\section*{Plain Language Summary}
Even in state-of-the-art climate models, the effects of many important small-scale processes cannot be directly simulated due to limited computing power. Thus, these effects are represented using functions called parameterizations. However, many of the current physics-based parameterizations have major shortcomings, leading to biases and uncertainties in the models’ predictions. Recently, there has been substantial interest in learning such parameterizations directly from short but very high-resolution simulations. Most studies have focused on using deep neural networks, which while leading to successful parameterizations in some cases, are hard to interpret and explain. A few more recent studies have focused on another class of machine-learning methods that discover equations. This approach has resulted in fully interpretable but unsuccessful parameterizations that produce unphysical results. Here, using widely-used test cases, we 1) explain the reasons for these unphysical results, 2) connect the discovered equations to well-known mathematically derived parameterizations, and 3) present ideas for learning successful parameterizations using equation-discovery methods. Our main finding is that the common loss functions that match patterns representing effects of small-scale processes are not enough, as important physical phenomena are not properly learned. Based on this, we have proposed a number of physics-aware metrics and loss functions for future work.


\section{Introduction} \label{sec: introduction}
Turbulent flows are ubiquitous in many geophysical systems, including atmospheric and oceanic circulations, and play an important role, e.g., greatly enhancing mixing and transport. Direct numerical simulation (DNS) of high-dimensional turbulent flows often becomes computationally intractable. Therefore, numerical simulations of most geophysical turbulent flows cannot resolve all the relevant scales \cite{fox2019challenges,palmer2001nonlinear, schneider2017climate}. Large-eddy simulation (LES) is a practical approach to balance computational cost and accuracy: the large scales of the flow are explicitly resolved, while the effects of the small-scale features which cannot be resolved by the given grid resolution, called subgrid-scale (SGS) features, are parameterized as a function of the resolved flow \cite{pope2000turbulent,sagaut2006large,smagorinsky1963general}. However, the performance of the LES models strongly depends on the accuracy of the employed SGS closure. Over years, there have been extensive efforts focused on formulating physics-based and semi-empirical SGS closures using various techniques in many turbulent flows \cite{meneveau2000scale,moser2021statistical,pope2000turbulent,sagaut2006large}, including geophysical flows \cite{alexander1999spectral,anstey2017deformation,berner2017stochastic,cessi2008energy,gallet2020vortex,herman2013linear,jansen2014parameterizing,o2008statistical,khodkar2019reduced,sadourny1985parameterization,schneider2017climate,sridhar2022large,sullivan1994subgrid,tan2018extended,zanna2017scale}.

The challenge of modeling SGS closures lies in faithfully representing the two-way interactions between the SGS processes and the resolved large-scale dynamics. There are two general approaches to SGS modeling: (a) functional and (b) structural \cite{sagaut2006large}. The functional SGS closures are developed by considering the inter-scale interactions (e.g., energy transfers). This is often achieved by introducing a dissipative term. Hence, functional SGS closures generally take an eddy-viscosity form to mimic the average function of the SGS eddies. Among the first and most-used functional closures is the Smagorinsky model  \cite{smagorinsky1963general}. Later, dynamic formulations of this model were proposed, in which the key coefficient is dynamically adjusted to the local structures of the flow \cite{germano1992turbulence,lilly1992proposed,ghosal1993local,chai2012dynamic}. Existing functional closures, most of which are the eddy-viscosity type, can be excessively dissipative \cite{vreman1996large, guan2022stable}. Furthermore, they cannot capture the structure of the SGS terms, leading to a low correlation coefficient (CC$<0.5$) with the true SGS terms, i.e.,  those diagnosed from the DNS data \cite{carati2001modelling,guan2022stable,moser2021statistical}. 

On the contrary, structural closures tend to have much higher CC with the true SGS terms. Structural closures approximate the SGS terms by constructing it from an evaluation of large-scale motions or a formal series expansion. One of the most common structural closures is the nonlinear gradient model \cite{leonard1975energy,clark1979evaluation}, referred to as NGM hereafter (it is also known as the tensor diffusivity model \cite{leonard1999tensor}). {\it The NGM can be derived analytically}: the SGS term is approximated using a first-order truncated Taylor-series expansion of the SGS stress' convolution integral (details discussed later). However, despite CC$>0.9$, LES with NGM closure has been found to be unstable in many studies of two-dimensional (2D) and three-dimensional (3D) turbulence. These instabilities are often attributed to insufficient dissipation and more importantly, to the presence of too-strong backscattering in NGM \cite{leonard1997large,leonard2016large,liu1994properties,fabre2011development,lu2010modulated,meneveau2000scale,prakash2021optimal,chen2003physical,chen2006physical,vollant2016dynamic, moser2021statistical}. As a result, while backscattering (basically anti-diffusion or up-gradient flux) is an important process to represent in closure models \cite{grooms2015numerical,guan2022stable,hewitt2020resolving,nadiga2010stochastic,shutts2005kinetic}, it is ignored in most practical SGS closures in favor of stability (though there has been some new exciting progress in this area; see, e.g., \citeA{jansen2015energy} and \citeA{juricke2020kinematic}). In fact, currently operational climate models do not account for backscattering in their ocean parameterizations \cite{hewitt2020resolving}. Consequently, a framework for developing SGS closures with the right amount of diffusion and backscattering, that can capture both the structure and function of the SGS terms, has remained elusive \cite{moser2021statistical, pope2000turbulent, sagaut2006large}. 

Before moving forward, it should be pointed out that while the discussion so far has been focused on closure for geophysical turbulence, many other critical processes in the Earth system (in atmosphere, ocean, land, cryosphere, biosphere and at their interfaces) require parameterizations in Earth system models \cite{stensrud2009parameterization,schneider2021accelerating}. Thus, the discussion below and as clarified later, the findings of this paper, are broadly relevant to parameterization efforts in Earth science.

Recently, machine learning (ML) has brought new tools into SGS closure modeling \cite{schneider2017earth,zanna2021deep,brunton2020machine,duraisamy2021perspectives,gentine2021deep, balaji2021climbing}. The strength of ML techniques is their ability to handle high-dimensional data and learn strongly nonlinear relationships. Therefore, ML techniques are attractive tools that might be able to extract more hidden knowledge from data, potentially providing better SGS closures and even new insights into SGS physics. Data-driven SGS closures, e.g., based on deep neural networks trained on high-fidelity simulation data such as DNS data, have been developed for canonical geophysical flows such as 2D and quasi-geostrophic turbulence \cite{bolton2019applications,frezat2022posteriori,guan2022stable,guan2023learning,pawar2020priori,maulik2018data, srinivasan2023turbulence} and oceanic and atmospheric circulations \cite{beucler2021enforcing,brenowitz2018prognostic,cheng2022deep,guillaumin2021stochastic,rasp2018deep,yuval2020stable,zhang2022seasonal}. While some of these studies found the learned data-driven SGS closures to lead to stable and accurate LES \cite{yuval2020stable,guan2022stable,guan2023learning,frezat2022posteriori}, a number of major challenges remain \cite{schneider2021accelerating, balaji2021climbing}. Perhaps the most important one is {\it interpretability}, which is difficult for neural networks, despite some recent advances in explainable ML for climate-related applications \cite{clare2022explainable,mamalakis2022investigating}, including for SGS modeling \cite{subel2022explaining,pahlavan2024explainable}. The black-box nature of neural network-based closures aside, there are also challenges related to generalizability, computational cost, and even implementation \cite{balaji2021climbing,chattopadhyay2020data,guan2022stable,kurz2020machine,maulik2019subgrid,subel2021data,xie2019artificial,zhou2019subgrid}, limiting the broad application of such closures in operational climate and weather models, at least for now. 

An alternative approach that is rapidly growing in popularity involves using ML techniques that provide interpretable, closed-form equations, e.g. using sparse linear regression. The underlying idea of this {\it equation-discovery} approach is that given spatial, temporal, or spatio-temporal data from a system, one can discover the governing (algebraic or differential) equations of that system \cite{brunton2016discovering,chen2022symbolic,goyal2022discovery,mojgani2022discovery,schneider2020imposing,rudy2017data,schaeffer2017learning,schmidt2009distilling,schneider2021learning,schneider2022ensemble,udrescu2020ai,zhang2018robust}. Most of the aforementioned studies are focused on discovering the entire governing equations from data, though few recent studies have used this approach to discover SGS closures (see below). This approach has the following advantages over more complex methods such as neural networks in the context of SGS modeling: 1) the learned closure is significantly easier to interpret based on physics \cite{zanna2020data}, 2) the number of required training samples and the training costs are often considerably lower \cite{brunton2020machine,mojgani2022discovery,mojgani2023interpretable}, and 3) the computational cost of implementation in conventional solvers is lower, as the discovered closures often involve traditional operations, e.g., gradients and Laplacians \cite{udrescu2020ai, ross2023benchmarking}.

A number of equation-discovery techniques and test cases have been recently employed for {\it structural} modeling of the SGS stress. In the first study of its kind, \citeA{zanna2020data} used relevance vector machine (RVM), a sparsity-promoting Bayesian linear regression technique, with a library of second-order velocity derivatives and their nonlinear combinations, to learn a closed-form closure model for the SGS momentum and buoyancy fluxes from {\it filtered} high-resolution simulations of ocean mesoscale turbulence. They found a closure that resembled the NGM, with close connections to earlier physics-based modeling work by \citeA{anstey2017deformation}. Although, the discovered closure performed well in \apriori (offline) tests, it was unstable \aposteriori (online), i.e., when it was coupled to a low-resolution ocean solver. Following the same general approach, more recently, \citeA{ross2023benchmarking} proposed a novel equation-discovery approach combining linear regression and \KJ{genetic programming}. This hybrid approach uses \KJ{genetic programming} to discover the structure of the equation followed by linear regression to fine-tune the coefficients. In contrast to methods such as RVM, \KJ{genetic programming} does not require an explicit library of features, instead, it uses a simple set of features and operations, and constructs expressions by successively applying operators and combining expressions. Similarly, in other disciplines, \citeA{reissmann2021application} and \citeA{li2021data} recently used \KJ{gene expression programming} to discover SGS stress for the Taylor-Green vortex and the 3D isotropic turbulence, respectively. They developed a nonlinear closure consisting of the local strain rate and rotation rate tensors, based on what is known as Pope tensors \cite{pope1975more}, which will be discussed later. Overall, these more recent studies found that \KJ{gene expression programming- and genetic programming-}based closures often outperform common baselines such as the Smagorinsky and the mixed models when turbulence statistics and flow structures are considered ~\cite{li2021data,reissmann2021application,ross2023benchmarking}. Note that there also have been a number of studies focused on equation-discovery for {\it functional} modeling, e.g., using techniques such as Ensemble Kalman inversion \cite{schneider2021learning, schneider2020imposing}; see the Summary and Discussion.  



In this study, we build on the work by \citeA{zanna2020data} and use 2D forced homogeneous isotropic turbulence (2D-FHIT) and Rayleigh-B\'enard convection (RBC) to extend and expand their analysis in several directions:
\begin{enumerate}
    \item We use RVM with an expansive high-order library to discover closures from DNS data for the SGS momentum flux tensor (2D-FHIT and RBC) and the SGS heat flux vector (RBC).
    \item We conduct extensive robustness analysis of the discovered closures across a variety of flow configurations, filter types, and filter sizes, and examine the potential effects of numerical errors.
    \item Further clarify the connections between the robustly discovered SGS momentum and heat flux closures, and the SGS closures obtained analytically from the truncated Taylor-series expansion of the filter's convolution integral, the NGM \cite{leonard1975energy}.
    \item Explain the physical reason for the unstable \aposteriori LES with the discovered SGS closures, despite their high \apriori accuracy in some metrics (such as CC).
    \item Present a decomposition of the SGS tensor to the Leonard, cross, and Reynolds components, showing their relative importance and dependence on the filter type/size. 
    \item Based on these findings, we present a number of ideas for discovering stable and accurate SGS closures from the data in future work.
\end{enumerate}
 Note that while we focus on the use of RVM here, our findings and conclusions in (1)-(6) are applicable to any equation-discovery effort, and not just for SGS momentum and heat fluxes in geophysical turbulence, but for SGS modeling in any nonlinear dynamical system. 
 
This paper is organized as follows. In \cref{sec: methods}, we provide an introduction to the methodology, including the governing equations of test cases (2D-FHIT and RBC), filtering procedure for data and equations, RVM algorithm, and the employed library of the basis functions. \cref{sec: results} includes the discussion on the discovered closures, \apriori and \aposteriori tests, connection with the physics-based closures, and contribution of the Leonard, cross, and Reynolds components. Summary and Discussion are in \cref{sec: summary and discussion}.


\section{Models, Methods, and Data} \label{sec: methods}
\subsection{Filtering Procedure}

In DNS, the velocity field, $\bm{u}(\bm{x},t)$, is resolved using high spatio-temporal resolutions down to all relevant scales. In LES, a low-pass filtering operation, denoted by $\filtered{(.)}$, is performed on the equations and flow fields. The resulting filtered fields, for example, filtered velocity, $\filtered{\bm{u}}\left(\bm{x},t\right)$, can be adequately resolved using relatively coarse spatio-temporal resolutions: the required grid spacing is proportional to the specified filter width, $\Delta$, which is analogous to the size of the smallest eddies resolved in the LES \cite{pope2000turbulent, sagaut2006large}. Using $\bm{u}(\bm{x},t)$ as an example, the general spatial filtering operation is defined by \cite{sagaut2006large}
\begin{equation} \label{eq:general filtering}
    \filtered{\bm{u}}(\bm{x},t) = G*\bm{u}\ = \int_{-\infty}^{\infty} G\left(\bm{r}\right) \bm{u}(\bm{x}-\bm{r},t) d\bm{r},
\end{equation}
where $*$ is the convolution operator, and the integration is performed over the entire domain. The specified filter kernel, $G$, satisfies the normalization condition
\begin{equation}
    \int_{-\infty}^{\infty} G\left(\bm{r}\right) d\bm{r} = 1.
\end{equation}
Subsequently, any flow field such as velocity can be decomposed into a filtered (resolved) part and SGS (residual) part:
\begin{equation}
    \bm{u}(\bm{x},t) = \filtered{\bm{u}}(\bm{x},t)+\bm{u}'(\bm{x},t),
\end{equation}
where $\bm{u}'$ is the SGS field. While this appears to be analogous to the Reynolds decomposition, an important distinction should be noted: the filtered residual field may not be strictly zero ($\filtered{\bm{u}'} \neq 0$, thus $\filtered{\filtered{\bm{u}}}\neq \filtered{\bm{u}}$), depending on the choice of the filter function \cite{sagaut2006large}. Further details about the filters used in this work (Gaussian, box, Gaussian + box, and sharp-spectral) are given in \ref{appendix: filtering procedure}.

\subsection{Two-dimensional Forced Homogeneous Isotropic Turbulence (2D-FHIT)}

 We consider 2D-FHIT as the first test case. This canonical flow has been extensively used for testing novel physics-based and ML-based SGS closures for geophysical turbulence in the past decades \cite{boffetta2012two,chandler2013invariant,guan2022stable,tabeling2002two,thuburn2014cascades, vallis2017atmospheric,verkley2019maximum}. The dimensionless continuity and momentum equations for 2D-FHIT in ($x,y$) spatial dimensions are:
\begin{eqnarray}
    \nabla \cdot \bm{u} & = & 0, \label{eq:2d-fhit uv continuity}\\
    \frac{\partial \bm{u}}{\partial t} + \left(\bm{u} \cdot \nabla\right) \bm{u} & = & 
    -\nabla p + 
    \frac{1}{Re} \nabla^2\bm{u}
    + \bm{\mathcal{F}} +\bm{\mathcal{R}}, \label{eq:2d-fhit uv momentum}
\end{eqnarray}
where $\bm{u} = \left(u,v\right)$ is the velocity, $p$ is the pressure, $\bm{\mathcal{F}}$ represents a time-constant external forcing, $\bm{\mathcal{R}}$ is the Rayleigh drag, and $Re$ is the Reynolds number. The domain is doubly periodic with length $L = 2\pi$.

The equations for LES are obtained by applying a homogeneous 2D filter (Eq. \eqref{eq:general filtering}) to Eqs.~\eqref{eq:2d-fhit uv continuity}-\eqref{eq:2d-fhit uv momentum}. The filtered continuity and momentum equations are:
\begin{eqnarray}
    \nabla \cdot \filtered{\bm{u}} & = & 0, \label{eq:2d-fhit filtered uv continuity}\\
    \frac{\partial \filtered{\bm{u}}}{\partial t} + \left(\filtered{\bm{u}} \cdot \nabla\right) \filtered{\bm{u}} & = & 
    -\nabla \,\filtered{p} + 
    \frac{1}{Re} \nabla^2\filtered{\bm{u}} - \nabla \cdot \bm{\uptau}
    + \filtered{\bm{\mathcal{F}}} +\filtered{\bm{\mathcal{R}}}, \label{eq:2d-fhit filtered uv momentum}
\end{eqnarray}
where $\bm{\uptau}$ is the SGS stress tensor:
\begin{eqnarray}
    \bm{\uptau} = 
    \begin{bmatrix}
    \tau_{xx} & \tau_{xy} \\
    \tau_{yx} & \tau_{yy}
    \end{bmatrix}
    =
    \begin{bmatrix}
    \,\filtered{u^2} - \filtered{u}^2 & 
    \,\filtered{uv} - \filtered{u}\,\filtered{v} \\
    \filtered{uv} - \filtered{u}\,\filtered{v} &
    \filtered{v^2} - \filtered{v}^2 & 
    \end{bmatrix}.
    \label{eq:2d-fhit sgs stress tensor}
\end{eqnarray}
A closure model is needed to represent $\tau_{xx}$, $\tau_{xy}$ $(=\tau_{yx})$, and $\tau_{yy}$, in terms of the resolved flow ($\filtered{u}, \filtered{v}, \filtered{p}$). However, currently, this is not possible just using the first principles due to the presence of the $\filtered{u^2}, \filtered{uv}, \text{ and } \filtered{v^2}$ terms.

We study three cases of 2D-FHIT (\cref{table:2d-fhit cases}), creating a variety of flows that differ in dominant length scales and energy/enstrophy cascade regimes. For DNS, as discussed in \ref{appendix: 2d-fhit numerical solver}, Eqs.~\eqref{eq:2d-fhit uv continuity}-\eqref{eq:2d-fhit uv momentum} are numerically solved at high spatio-temporal resolutions using a Fourier-Fourier pseudo-spectral solver. For the LES, the same solver at lower spatio-temporal resolution is used (\ref{appendix: 2d-fhit numerical solver}).

\begin{table}
\centering
\caption {Physical and numerical parameters used in the 2D-FHIT cases. Cases with different flow regimes are produced by varying forcing wavenumber,  $\left(f_{k_{x}},f_{k_{y}}\right)$, and $Re$. \KJ{The spatial scales of the dominant flow structures depend on forcing wavenumber; the higher the forcing wavenumber, the smaller the scales. Cases K1 and K3 exhibit a dominance of large-scale structures\KJnew{, while small-scale structures prevail in Case K2. See \citeA{guan2023learning} for further discussions of the dynamical differences between these cases}. The increase in $Re$ adds more small-scale features in $\omega$, and changes the spectrum of the SGS stress tensor, $\bm{\uptau}$, in both large and small scales}. For each case, we use several filter types (Gaussian, box, Gaussian $+$ box, and sharp-spectral filters) and filter sizes, \KJ{$\Delta = 2\Delta_{\text{LES}}$, where $\Delta_{\text{LES}}={L}/{N_{\text{LES}}}$ is the LES grid spacing}. Here, $N_{\text{LES}} = \{32, 64, 128, 256\}$ for Cases K1 and K3 and $N_{\text{LES}} = \{128, 256\}$ for Case K2. Here, $N_{\text{LES}}$ and $N_{\text{DNS}}$ are the number of points in each direction on the LES and DNS grids, respectively. $L = 2\pi$ is the length of the domain. Note that the lowest $N_{\text{LES}}$ is chosen such that the LES resolution resolves at least $80\%$ of the DNS kinetic energy \cite{pope2000turbulent}. Filters are applied in both spatial dimensions for 2D-FHIT.}
\begin{tabular}{ |c|c|c|c| } 
\hline
Cases & $Re$      & ($f_{k_x}, f_{k_y}$)  & $N_{\text{DNS}}$ \\
\hline \hline
K1   & 20,000  & $\left(4,0\right)$        & $1024$ \\
K2   & 20,000  & $\left(25,25\right)$      & $1024$ \\
K3   & 100,000 & $\left(4,0\right)$        & $2048$ \\
\hline
\end{tabular}
\label{table:2d-fhit cases}
\end{table}

\subsection{Turbulent Rayleigh-B\'enard Convection (RBC)}
As our second test case, we use 2D turbulent RBC, a widely used canonical flow for buoyancy-driven turbulence \cite{chilla2012new,dabbagh2017priori,hassanzadeh2014wall,kooloth2021coherent,lappa2009thermal,sondak2015optimal}, which in addition to the SGS (momentum) stress, requires closure modeling of the SGS heat flux \cite{pandey2022direct,peng2002subgrid,wang2008new}. Under the Oberbeck-Boussinesq approximation, the dimensionless governing equations for the flow between horizontal walls at fixed temperatures (the bottom wall being warmer than the top) in ($x,z$) spatial dimensions are:

\begin{eqnarray}
    \nabla \cdot \bm{v} & = & 0,\label{eq:rbc uv continuity} \\
    \frac{\partial \bm{v}}{\partial t} + \bm{v} \cdot \nabla \bm{v} & = & -\nabla p + Pr \nabla^2\bm{v} +  Ra \, Pr \,\theta \hat{\bm{z}}, \label{eq:rbc uv momentum} \\
    \frac{\partial \theta}{\partial t} + \bm{v} \cdot \nabla \theta - w& = &   \nabla^2 \theta, \label{eq:rbc uv energy}
\end{eqnarray}
where $\bm{v}=\left(u,w\right)$ is the velocity, $\theta$ is the temperature ($T$) departure from the conduction state, $\hat{\bm{z}}$ is the unit vector in the vertical direction, and $Ra$ and $Pr$ are the Rayleigh and Prandtl numbers, respectively. The domain is periodic in the horizontal direction with horizontal domain size $L=6\pi$ \KJ{and vertical length of 1}. No-slip boundary conditions are applied at the walls. We use three cases of turbulent RBC (\cref{table:rbc cases}) in which the $Ra$ and $Pr$ are varied. 

To properly resolve the thin boundary layers in turbulent RBC, a pseudo-spectral solver with (non-uniform) Chebyshev collocation points in the vertical direction is used. However, filtering variables on a non-uniform grid can cause major errors in the diagnosed SGS terms, because the filters will not commute with spatial derivatives \cite{yalla2021effects}. As a result, following the common practice for LES, we only filter the equations in the horizontal direction, where (uniform) Fourier collocation points are used. The LES equations obtained by applying a 1D filter along the horizontal direction, $x$, to Eqs.~\eqref{eq:rbc uv continuity}-\eqref{eq:rbc uv energy} are:
\begin{eqnarray}
    \nabla \cdot \filtered{\bm{v}} & = & 0,\label{eq:rbc filtered uv continuity} \\
    \frac{\partial \filtered{\bm{v}}}{\partial t} + \filtered{\bm{v}} \cdot \nabla \,\filtered{\bm{v}} & = & -\nabla \, \filtered{p} + Pr\, \nabla^2\filtered{\bm{v}} +  Pr\,Ra\,\filtered{\theta} \hat{\bm{z}} - \nabla \cdot {\bm{\uptau}}, \label{eq:rbc filtered uv momentum} \\
    \frac{\partial \filtered{\theta}}{\partial t} + \filtered{\bm{v}} \cdot \nabla \filtered{\theta} -\filtered{w} & = & \nabla^2 \filtered{\theta} - \nabla \cdot {\bm{J}}, \label{eq:rbc filtered uv energy}
\end{eqnarray}
where $\bm{\uptau}$ is the SGS (momentum) stress tensor
\begin{eqnarray}
    \bm{\uptau} = 
    \begin{bmatrix}
    \tau_{xx} & \tau_{xz} \\
    \tau_{zx} & \tau_{zz}
    \end{bmatrix}
    =
    \begin{bmatrix}
    \,\filtered{u^2} - \filtered{u}^2 & 
    \,\filtered{uw} - \filtered{u}\,\filtered{w} \\
    \filtered{uw} - \filtered{u}\,\filtered{w} &
    \filtered{w^2} - \filtered{w}^2 & 
    \end{bmatrix},
    \label{eq:rbc sgs stress tensor}
\end{eqnarray}
and $\bm{J}$ is the SGS heat flux vector
\begin{eqnarray}
    \bm{J} = \begin{bmatrix}
    J_x \\ J_z
    \end{bmatrix} = 
    \begin{bmatrix}
    \, \filtered{u\theta} - \filtered{u} \, \filtered{\theta} \,\\
    \, \filtered{w\theta} - \filtered{w} \, \filtered{\theta} \,
    \end{bmatrix}.\label{eq:rbc sgs heat flux}
\end{eqnarray}
Here, in addition to $\bm{\uptau}$, $\bm{J}$ needs a closure model too.

For DNS, as discussed in \ref{appendix: rbc numerical solver}, Eqs.~\eqref{eq:rbc uv continuity}-\eqref{eq:rbc uv energy} are numerically solved at high spatio-temporal resolutions using a Fourier-Chebyshev pseudo-spectral solver. For LES, the same solver with lower spatial resolution is used (\ref{appendix: rbc numerical solver}).

\begin{table}
\centering
\caption {Physical and numerical parameters used in three cases of turbulent RBC. Cases with different flow regimes are produced by varying $Ra$ and $Pr$. \KJ{Increasing $Ra$ enhances heat transfer rate and makes the flow structures more complex, where large-scale plumes may break down into smaller structures. $Pr$ increases the heat transfer efficiency and results in more small-scale plumes.} For each case, we use several filter types (Gaussian, box, Gaussian $+$ box, and sharp-spectral cutoff filters) and filter size \KJ{$\Delta= \Delta_x = 2\Delta_{\text{LES}}$, where $\Delta_{\text{LES}} = {L}/{N_{\text{LES}}}$ is the LES grid spacing.} Here, $N_{\text{LES}} = \{128, 256\}$ for Case R1 and $N_{\text{LES}} = \{128, 256, 512\}$ for Cases R2 and R3. Here, $N_{\text{LES}}$ is the number of points on the LES grid in the horizontal direction, $x$. $N^{\text{DNS}}_x$ and $N^{\text{DNS}}_z$ are the number of grid points on the DNS grid in the horizontal and vertical directions, respectively. $L = 6 \pi$ is the length of the domain in the horizontal direction. Note that the lowest $N_{\text{LES}}$ is chosen such that the LES resolution resolves at least $80\%$ of the DNS kinetic energy \cite{pope2000turbulent}. Filters are only applied along the horizontal direction. } 
\begin{tabular}{ |c|c|c|c| } 
\hline
Cases & $Ra$            & $Pr$  & $\left(N^{\text{DNS}}_x, N^{\text{DNS}}_z\right)$        \\
\hline \hline
R1    & $10^6$          & 100 & $\left(2048, 400\right)$ \\
R2    & $40 \times 10^6$ & 7   & $\left(2048, 400\right)$ \\
R3    & $40 \times 10^6$ & 100 & $\left(2048, 400\right)$ \\
\hline
\end{tabular}
\label{table:rbc cases}
\end{table}

\subsection{Filtered Direct Numerical Simulation (FDNS) Data}
It should be highlighted that in this study with two canonical test cases, we consider DNS data as the ``truth'', and use filtered DNS (FDNS) data to discover the closures. However, in reality, performing DNS for many geophysical flows is computationally prohibitive. In such cases, high-resolution LES that adequately resolves the process of interest (e.g., ocean eddies, gravity waves, etc.) is often used as the truth to train the ML algorithms for SGS modeling~\cite{yuval2020stable, zanna2021deep, shen2022library,sun2023quantifying}. 

Here, we compute FDNS variables on the LES grids, which are 4 to 64 times coarser than the DNS grid in both spatial dimensions for 2D-FHIT and one spatial dimension for RBC (see Tables \ref{table:2d-fhit cases}-\ref{table:rbc cases}). More specifically, we first apply the respective filter's transfer function (\cref{table:1d filters,table:2d filters}) to the DNS data, and then coarse-grain the results onto the LES grid. Note that following some of the recent papers \cite{grooms2021diffusion,guan2022stable}, we define ``filtering'' as an operation that removes the small scales but keeps the grid resolution (e.g., DNS), and ``coarse-graining'' as an operation that changes the grid size, e.g., from the DNS resolution to LES resolution. Note that $\bm{\uptau}$ and $\bm{J}$ in Eqs.~\eqref{eq:2d-fhit filtered uv momentum}, \eqref{eq:rbc filtered uv momentum}, and \eqref{eq:rbc filtered uv energy} need to be on the LES grid.

The filtering and coarse-graining are performed following \citeA{sagaut2006large} and \citeA{guan2022stable}. Briefly, using the velocity $\bm{u}\left({\bm{x}_{\text{DNS}}},t\right)$ as an example, and denoting the DNS grid and wavenumber as $\bm{x}_{\text{DNS}}$ and $\bm{k}_{\text{DNS}}$, we first transform the DNS velocity into the spectral space $\hat{\bm{u}}\left(\bm{k}_{\text{DNS}},t\right)$, where $\left(\hat{.}\right)$ means Fourier transformed. This is followed by applying the filter in the spectral space:
\begin{eqnarray} \label{eq:filtered dns spectral velocity}
    \filtered{\hat{\bm{u}}}\left(\bm{k}_{\text{DNS}},t\right) = \hat{G}\left(\bm{k}_{\text{DNS}}\right) \odot \hat{\bm{u}}\left(\bm{k}_{\text{DNS}},t\right).
\end{eqnarray}
Here, $\hat{G}\left(\bm{k}_{\text{DNS}}\right)$ can be any of the transfer functions listed in \cref{table:1d filters,table:2d filters}, and $\odot$ is the Hadamard (element-wise) multiplication. After the filtering operation, coarse-graining is performed to transform the filtered variable from the DNS to the LES grid. In this study, we perform coarse-graining in spectral space with cutoff, \KJ{$k_c=\pi/\Delta_{\text{LES}}$}, which for example in 2D, yields
\begin{eqnarray} \label{eq:filtered les 2d spectral velocity}
    \filtered{\hat{\bm{u}}}(\bm{k}_{\text{LES}},t) = \filtered{\hat{\bm{u}}}\left(\left|k_{\text{DNS},x}\right|<k_c, \left|k_{\text{DNS},y}\right|<k_c,t\right).
\end{eqnarray}

\KJ{It is important to highlight that the filtering operation can be reversible under certain conditions: The DNS data can be recovered (via deconvolution) from the filtered data (still on the DNS grid) if the filter is a Reynolds operation, not a projection \cite{sagaut2006large}. This is the case for Gaussian, box, and Gaussian + box filters, whose transfer function only attenuates the signal. However, for the sharp-spectral cutoff filter (which is a projection operator), and for coarse-graining, information is lost beyond the cutoff wavenumber \(k_c\) and therefore data recovery is limited only up to \(k_c\). This is further discussed in \cref{sec:gm}.}

Hereafter, for brevity, we use the term ``filtered'' (still denoted by $\overline{\cdot}$) to mean ``filtered'' and then ``coarse-grained''. \KJ{The spectrum of a filtered and coarse-grained variable can be found in \Cref{fig: 1D transfer function}(b) in \ref{appendix: filtering procedure}.}

\Cref{fig: snapshot 2dfhit rbc flow} shows the effects of filtering on the vorticity and temperature fields for 2D-FHIT and RBC, illustrating that the small-scale structures of $\omega$ and $T$ are removed due to filtering and the fields are smoothed out.

\begin{figure}
  \centering
  \includegraphics[width=\textwidth]{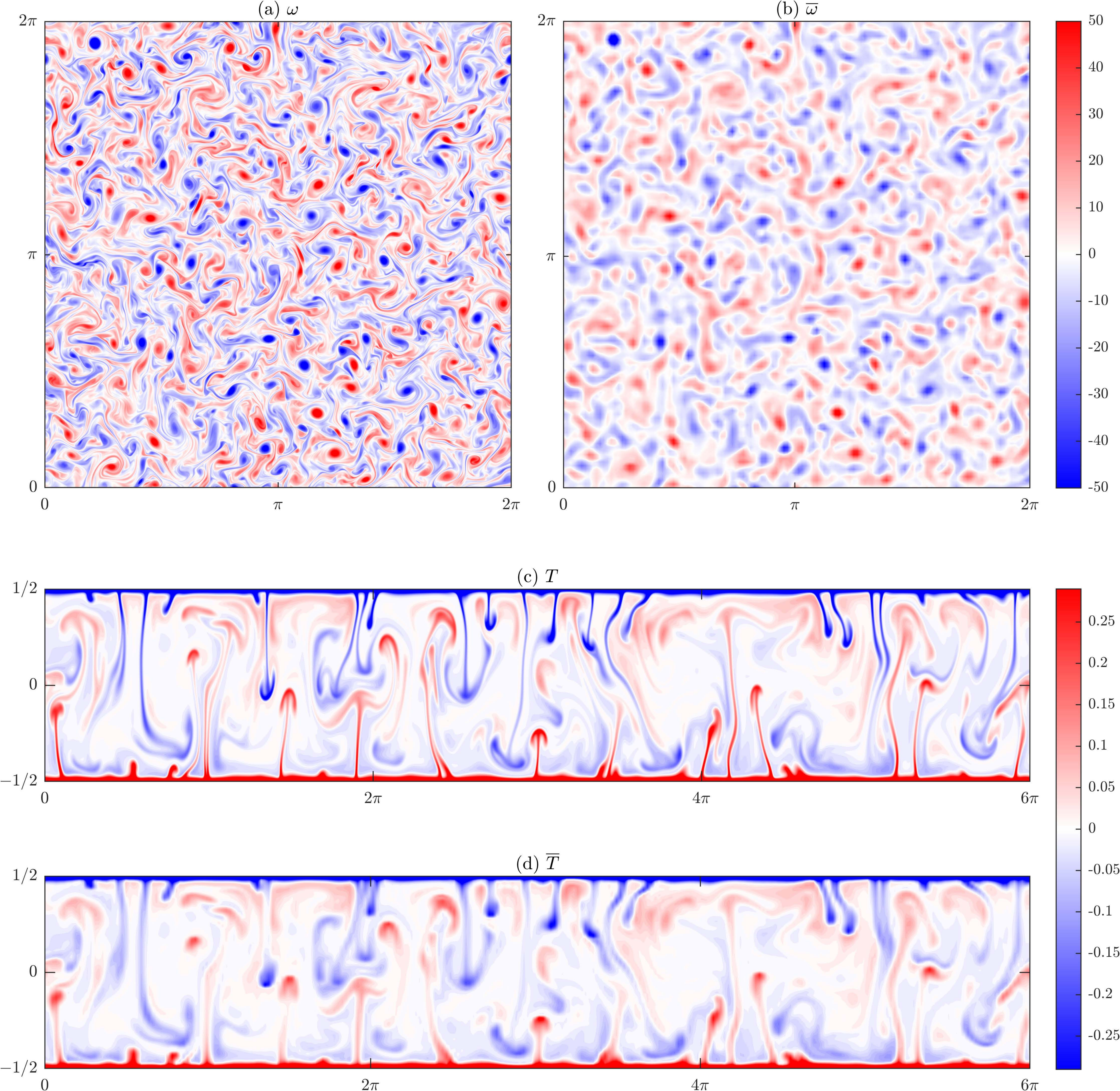}
  \caption{Snapshots of the (a) DNS vorticity field $\omega$ ($N_{\text{DNS}}=1024$) and the (b) FDNS vorticity field $\overline{\omega}$ ($N_{\text{LES}}=128$) for Case K2 (see \cref{table:2d-fhit cases}). The (c) DNS temperature field $T$ ($N^{\text{DNS}}_x=2048$), and the (d) FDNS temperature field ${\overline{T}}$ ($N_{\text{LES}}=256$) for Case R3 (see \cref{table:rbc cases}). The Gaussian filter is applied in both cases.}
  \label{fig: snapshot 2dfhit rbc flow}
\end{figure}

\subsection{The Equation-discovery Method}
In this study, we employ the RVM \cite{tipping2001sparse} to discover closed-form closures for each element of the $\bm{\uptau}$ tensor and $\bm{J}$ vector from the FDNS data. RVM is a sparsity-promoting Bayesian (linear) regression technique that has shown promise in applications involving dynamical systems \cite{zhang2018robust,zanna2020data,mojgani2022discovery}. RVM relies on a pre-specified library of basis functions $\bm{\Phi}$; each column of this matrix is a basis, e.g., a linear or nonlinear combination of relevant variables such as velocity and temperature and/or their derivatives. The library should be expressive enough so that $\bm{{s}}$, a vectorized snapshot of a element of any $\bm{\uptau}$ or $\bm{J}$, could be completely represented as
\begin{eqnarray}
    \bm{s}^{\text{RVM}} = \bm{\Phi}\bm{c}.
\end{eqnarray}
The vector of regression weights, $\bm{c}$, is computed by minimizing the mean-squared error (MSE)
\begin{eqnarray}
    \text{MSE}=\lVert \mathcal{S}^{\text{RVM}} - \mathcal{S}^{\text{FDNS}} \rVert^2_2, 
    \label{eq:mse}
\end{eqnarray}
where vector $\mathcal{S}$ consists of $n$ samples of $\bm{s}$ stacked together. \KJ{RVM assumes Gaussian prior distributions for each weight, and the width of the Gaussian posterior provides a measure of the weight's uncertainty. Sparsity is enforced via an iterative process: basis functions whose weights' uncertainties exceed a pre-specified hyperparameter (threshold), $\alpha$, are removed (pruned), and Eq.~\eqref{eq:mse} is minimized again. The iterations stop when all the remaining basis functions have uncertainties smaller than $\alpha$. Larger $\alpha$ results in a lower MSE but more terms in the discovered model (see below)}.

A critical step in using RVM (and most equation-discovery methods) is the choice of the library. Here, we have chosen the following libraries. For momentum stress, we use
\begin{eqnarray}
    \left[ \frac{\partial^{\left(q_1+q_2\right)} A}{\partial x^{q_1} \partial y^{q_2}}\right]^{p_1}
    \left[ \frac{\partial^{\left(q_4 + q_5\right)} B}{\partial x^{q_4} \partial y^{q_5}}\right]^{p_2} \; \; \mathrm{or} \; \; \;
    \left[\frac{\partial^{\left(q_1+q_2\right)} C}{\partial x^{q_1} \partial z^{q_2}}\right]^{p_1}
    \left[\frac{\partial^{\left(q_4 + q_5\right)} D}{\partial x^{q_4} \partial z^{q_5}}\right]^{p_2}; \label{eq:lib1}
\end{eqnarray}
where $A,B=\filtered{u}$ or $\filtered{v}$ (2D-FHIT) and $C,D=\filtered{u}$ or $\filtered{w}$ (RBC). Note that experiments with including $\filtered{\theta}$ in $D$ yield the same results. For heat flux, we use
\begin{eqnarray}
    \left[ \frac{\partial^{\left(q_1+q_2\right)} A}{\partial x^{q_1} \partial z^{q_2}}\right]^{p_1}
    \left[ \frac{\partial^{\left(q_4 + q_5\right)} \filtered{\theta}}{\partial x^{q_4} \partial z^{q_5}}\right]^{p_2}, \label{eq:lib2}
\end{eqnarray}
where $A=\filtered{u}, \filtered{w}$, or $\filtered{\theta}$ (RBC). These libraries are expansive, with integers $0 \le q \le 8$ and $0 \le p \le 2$, though the total derivative order is limited to 8th (there are a total of 546 and 614 terms in the libraries used for momentum and heat fluxes, respectively). The form of these libraries is motivated by the Galilean-invariant property of the SGS terms, and by past studies. For example, these libraries include Pope's tensors \cite{pope1975more}, which have been used in physics-based \cite{anstey2017deformation,gatski1993explicit,jongen1998general,lund1993parameterization} and equation-discovery \cite{li2021data,reissmann2021application, ross2023benchmarking} approaches in the past (and include the structure of the Smagorinsky model; see below). Our library also includes the basis functions used by \citeA{zanna2020data}. 

Note that all calculations for the libraries (and any computation in this work) is performed using the same spectral methods used for DNS and LES.

We have found it useful for interpretability of the outcome and improving the robustness of the algorithm to remove redundant terms using the continuity equation (e.g., using $\partial \filtered{v}/{\partial y}=-\partial \filtered{u}/{\partial x}, \partial^2 \filtered{v}/{\partial y\partial x}=-\partial^2 \filtered{u}/{\partial x^2}$, etc.). Also, we have found it essential to normalize each basis in $\bm{\Phi}$ to have a zero mean and a unit variance, because the amplitude of higher-order derivatives can be much larger than that of the lower-order ones. 

Like any method, equation discovery using RVM has a number of strengths and weaknesses:
\begin{enumerate}
    \item It is data efficient \cite{zanna2020data, mojgani2022discovery}. For example, here, we report the results with $n=100$ FDNS samples, but even with $n=1$, the results remain practically the same.
    \item It is more robust, in terms of convergence, compared to similar sparsity-promoting techniques \cite{zhang2018robust, zanna2020data}.
    \item A pre-specified library is needed and it is assumed that the true answer (e.g., the SGS stress) can be represented as a linear combination of the chosen basis functions.
    \item The pre-specified hyper-parameter $\alpha$ determines how parsimonious the discovered model is. Decreasing $\alpha$ leads to a smaller (likely, more interpretable) model at the expense of increasing the MSE. Here, we follow the model-selection literature \cite{mangan2017model, mojgani2022discovery} and objectively choose $\alpha$ using the L-curve, as shown later.
    \item The answer can depend on the choice of the loss function. The RVM's MSE loss (Eq.~\eqref{eq:mse}) is strictly following the principle of structural modeling,  matching the flux between the FDNS and the discovered model. 
\end{enumerate}
Note that the above strengths (1)-(2) are highly desirable while these weaknesses (3)-(5) are common among many equation-discovery methods, although techniques such as \KJ{genetic programming} and \KJ{gene expression programming} can address (3) and (5), for example using an evolving library. We will further discuss (3)-(5) in Section~\ref{sec: summary and discussion}.





\section{Results} \label{sec: results}
In this section, we present and discuss the discovered closures, and analyze them \apriori (offline) and \aposteriori (online, coupled with LES).  We then uncover the connections between the discovered closure and the NGM. For all results presented here, we use $n=100$ FDNS samples from a training set and $20$ FDNS samples from an independent testing set.

\subsection{The Discovered Closures for SGS Momentum and Heat Fluxes} \label{subsec:discovered model}

For each of the six cases in Tables~\ref{table:2d-fhit cases}-\ref{table:rbc cases}, we separately discover closures for three elements of the SGS stress tensor, i.e., $\tau_{xx}, \tau_{xy}=\tau_{yx}$, and $\tau_{yy}$ for 2D-FHIT, and $\tau_{xx}, \tau_{xz}=\tau_{zx}$, and $\tau_{zz}$ for RBC. Additionally, we discover two elements of the SGS heat flux vector, i.e., $J_x$ and $J_z$ for RBC. We discover individual closures for 4 filter types: Gaussian, box, sharp-spectral, and Gaussian + box. The first three are common filter types, while the last one is motivated by a few recent studies~\cite{zanna2020data,guillaumin2021stochastic}. We also examine several filter sizes, $\Delta$ (see Tables~\ref{table:2d-fhit cases}-\ref{table:rbc cases}), and the effect of varying $\alpha$, which as mentioned earlier, is a key hyper-parameter in RVM.  

We analyze the \apriori performance of the discovered closures using the most commonly used metric: the average of CCs for testing samples \cite{sagaut2006large,maulik2019subgrid, guan2023learning}. For each element of $\bm{\uptau}$ or $\bm{J}$, denoted below by $\tau$ for convenience, the CC for each testing sample is calculated between 2D patterns of $\tau$ from FDNS and $\tau$ predicted by the RVM-discovered closure for the corresponding filtered flow variables (e.g., $\overline{u}$, $\overline{v}$ etc.): 
\begin{eqnarray}
    \text{CC} = \frac{\langle \left( \tau^{\text{RVM}} - \langle \tau^{\text{RVM}} \rangle \right) \left( \tau^{\text{FDNS}} - \langle \tau^{\text{FDNS}} \rangle \right)\rangle}{\sqrt{\langle \left(\tau^{\text{RVM}} - \langle \tau^{\text{RVM}}\rangle \right)^2 \rangle}\sqrt{\langle \left(\tau^{\text{FDNS}} - \langle \tau^{\text{FDNS}} \rangle \right)^2 \rangle}},
\end{eqnarray}
where $\langle\cdot\rangle$ is domain averaging. The same equation is also used for computing CC values of 2D patterns of inter-scale energy or enstrophy transfer, $P$ (defined later). 

\begin{figure}[t]
  \centering
  \includegraphics[width=\textwidth]{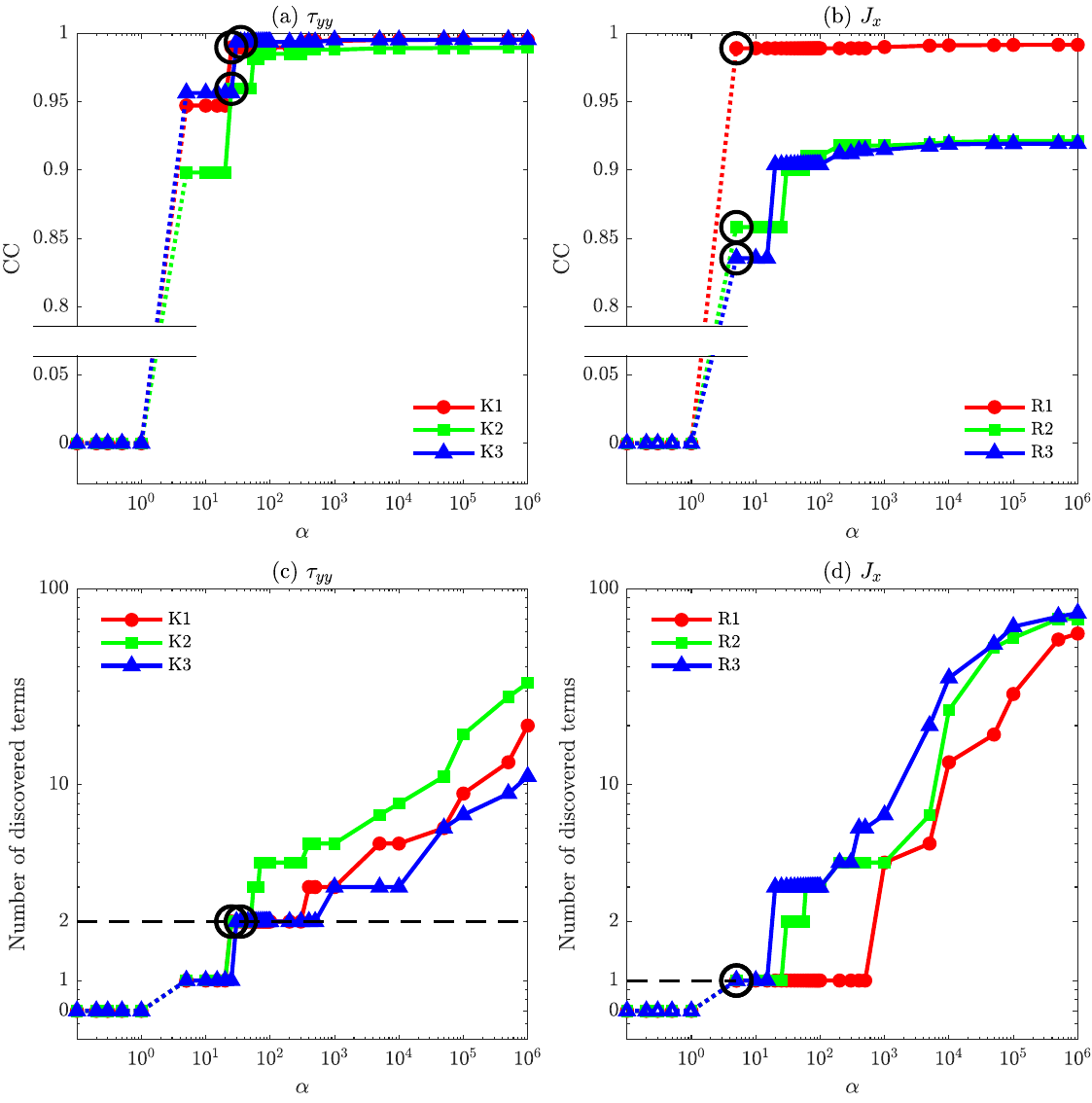}
  \caption{Representative examples of the effects of increasing the sparsity-level hyper-parameter, $\alpha$, on the CC and number of terms in the discovered closure. (a), (c): $\tau_{yy}$ (2D-FHIT) and (b), (d): $J_x$ (RBC). \KJ{A Gaussian filter with $N_{\text{LES}}=128$ (for Cases K1-K3) and $N_{\text{LES}}=256$ (for Cases R1-R3) is used}, but the same behavior is observed with any other $N_{\text{LES}}$ and filter type (except for the sharp-spectral, see the text). In general, for small $\alpha \; (<1)$, no closure is discovered (CC=0, zero term). With increasing $\alpha$, the CC converges to $\sim 1$ (a more accurate \apriori closure) but at the expense of a larger closure with many more terms (note the logarithmic scale of the $y$ axes in panels (c)-(d)). However, the CC-$\alpha$ relationship forms an ``L-curve'', whose elbow indicates the $\alpha$ that balances accuracy and model size (see the text).
 }
  \label{fig: CC number of discovered terms}
\end{figure}

As a representative example of the findings, \cref{fig: CC number of discovered terms}(a)-(b) shows the averaged CC for $\tau_{yy}$ (K1-K3) and $J_x$ (R1-R3) as $\alpha$ is increased. \Cref{fig: CC number of discovered terms}(c)-(d) presents the number of terms in the discovered closures. With small $\alpha$, the discovery is unsuccessful (CC=0; zero terms). However, as $\alpha$ is further increased, for all cases, CC abruptly jumps to above $0.8-0.9$ with 1-2 discovered terms, and then gradually converges to $1$ but with exponentially growing number of terms in the discovered closure. The CC-$\alpha$ relationship forms an ``L-curve''. The elbow of this curve indicates the $\alpha$ that balances accuracy and model size, and is extensively used in the model-selection and equation-discovery literature to objectively choose $\alpha$ \cite{Lawson_1995, Calvetti_JCAM_2000,mangan2017model,goyal2022discovery,mojgani2022discovery}. Examining all cases with other filter sizes and filter types reveals the same behavior as shown in \cref{fig: CC number of discovered terms}, with the exception of the sharp-spectral filter \KJ{as shown in \cref{fig: CC number of discovered terms sharp spectral}}. For this filter, the discovery is unsuccessful, leading to low CC and non-robust results; we will explain the reason for this failure later in this section.

We use the L-curve to determine the {\it optimal} $\alpha$. In 2D-FHIT, there are two kinks in the curve around the elbow, corresponding to the discovery of closures with 1 and 2 terms, respectively (\cref{fig: CC number of discovered terms}(a), (c)). Given the robust and asymptotic behavior in $\alpha$ after the second kink, we use the corresponding $\alpha$ to identify the discovered closure (see the black circles). We find that consistently, across Cases K1-K3, filter types, and filter sizes, this closure is of the form  
\begin{eqnarray}
    \bm{\uptau} = \begin{bmatrix}
        \tau_{xx} & \tau_{xy} \\
        \tau_{yx} & \tau_{yy}
    \end{bmatrix}
    = \Delta^2
    \begin{bmatrix}
        a_{xx} \left(\frac{\partial \filtered{u}}{\partial x}\right)^2 + b_{xx} \left(\frac{\partial \filtered{u}}{\partial y}\right)^2 &
        a_{xy} \frac{\partial \filtered{u}}{\partial x} \frac{\partial \filtered{v}}{\partial x} +
        b_{xy} \frac{\partial \filtered{u}}{\partial y} \frac{\partial \filtered{v}}{\partial y} \\ 
        a_{xy} \frac{\partial \filtered{u}}{\partial x} \frac{\partial \filtered{v}}{\partial x} +
        b_{xy} \frac{\partial \filtered{u}}{\partial y} \frac{\partial \filtered{v}}{\partial y}  &
        a_{yy} \left(\frac{\partial \filtered{v}}{\partial x}\right)^2 + b_{yy} \left(\frac{\partial \filtered{v}}{\partial y}\right)^2 
    \end{bmatrix}, 
    \label{eq:2d-fhit tau discovered model}
\end{eqnarray}
where $a_{xx}$, $b_{xx}$, $a_{xy}$, $b_{xy}$, $a_{yy}$, and $b_{yy}$ are the discovered coefficients ($\Delta^2$ is factored out to further highlight the independence of these coefficients from the filter size). Table~\ref{table:2d-fhit tau coeff} shows that these 6 coefficients are the same, and the same for Cases K1-K3, although they can depend on the filter type. This table also shows the average CC values of the discovered closure, which are around $0.99$, demonstrating the accurate prediction of each element of the stress tensor and the excellent \apriori (offline) performance of the discovered closure for a broad range of LES resolutions. 

Following the same approach, we discover basically the same closure for $\bm{\uptau}$ in RBC
\begin{eqnarray}
    \bm{\uptau} = \begin{bmatrix}
        \tau_{xx} & \tau_{xz} \\
        \tau_{zx} & \tau_{zz}
    \end{bmatrix}
    = \Delta^2
    \begin{bmatrix}
        d_{xx} \left(\frac{\partial \filtered{u}}{\partial x}\right)^2 &
        d_{xz} \frac{\partial \filtered{u}}{\partial x} \frac{\partial \filtered{w}}{\partial x} \\ 
        d_{xz} \frac{\partial \filtered{u}}{\partial x} \frac{\partial \filtered{w}}{\partial x} &
        d_{zz} \left(\frac{\partial \filtered{w}}{\partial x}\right)^2
    \end{bmatrix}, \label{eq:rbc tau discovered model}
\end{eqnarray}
where, as before, $d_{xx}$, $d_{xz}$, and $d_{zz}$ are the coefficients with $\Delta^2$ factored out. Note that Eq.~\eqref{eq:rbc tau discovered model} is the same as Eq.~\eqref{eq:2d-fhit tau discovered model}, except that here, there is one term rather than two in each element of the tensor, which is a result of filtering (in RBC) performed only in the horizontal, $x$, direction. As before, Table~\ref{table:rbc tau coeff} shows that these $d$ coefficients are the same, and the same for Cases R1-R3, though varying with filter type. Like before, the discovered closure has fairly high CC values. 

Again, following the same approach, we determine the optimal $\alpha$ for discovering the closure of $\bm{J}$. In \cref{fig: CC number of discovered terms}(b), Case R1 has a clear elbow while Cases R2-R3 have two kinks around the elbow. Examining all cases and the number of discovered terms (\cref{fig: CC number of discovered terms}(d)), we find that the single-term closures discovered at the first kink (circled) provide consistent and robust results. \KJ{The first kink is observed for all the Cases R1-R3. In contrast, the second kink is observed only in Cases R2-R3. The closure at the first kink is}
\begin{eqnarray}
    \bm{J}=\begin{bmatrix} 
        J_x \\
        J_z
    \end{bmatrix}
    = \Delta^2
    \begin{bmatrix}  
        d_x \frac{\strut \partial \filtered{u}}{\strut \partial x} \frac{\strut \partial \filtered{\theta}}{\strut \partial x} \\
        d_z \frac{\strut \partial \filtered{w}}{\strut \partial x} \frac{\strut \partial \filtered{\theta}}{\strut \partial x}
    \end{bmatrix},\label{eq:rbc J discovered model}
\end{eqnarray}
where $d_x$ and $d_z$ are the discovered coefficients with $\Delta^2$ factored out. Table~\ref{table:rbc j coeff} shows that these $d$ coefficients are the same, and the same for Cases R1-R3, but varying with filter type. As before, the discovered closure has a good \apriori performance.


To summarize the findings, Eqs.~\eqref{eq:2d-fhit tau discovered model}-\eqref{eq:rbc J discovered model} and Tables~\ref{table:2d-fhit tau coeff}-\ref{table:rbc j coeff} show that
\begin{enumerate}
\item Closures of the same form are robustly discovered for $\bm{\uptau}$ in two vastly different systems, 2D-FHIT and RBC. Even the closure for $\bm{J}$ overall has the same form, consisting of the products of the first-order derivatives of the variables involved in the nonlinearity of the SGS term.
\item Not just the form, but even the coefficients of the terms in the closures, are consistently the same as parameters such as $Re$, forcing wavenumber, $Ra$, or $Pr$ are changed in Cases K1-K3 and R1-R3, leading to different dynamics. The coefficients are independent of the {\it fluid} and even the {\it flow} properties.
\item The form of the closures is independent of the filter type unless the sharp-spectral filter is used. The coefficients, once normalized by $\Delta^2$, are independent of filter size, but depend on filter type.  
\item The discovered closures have outstanding \apriori performance, often with CC$ >0.95$ and even as high as $0.99$. It should be noted that the CCs reported in these tables are averaged over a broad range of $N_{\rm{LES}}$. The values of CC are higher for larger $N_{\rm{LES}}$, i.e., smaller $\Delta$.    
\end{enumerate}

\begin{landscape}
\begin{table} [tp]
\centering
\caption {Coefficients in Eq.~\eqref{eq:2d-fhit tau discovered model}, the robustly discovered closure for $\bm{\uptau}$ for 2D-FHIT (note that $\Delta^2$ is included in the coefficients). For Cases K1-K3 and different filter types, the mean and standard deviation of the discovered coefficients over different $N_{\text{LES}}$ are reported (see Tables~\ref{table:2d-fhit cases}). The average CC of the closure for each element of $\bm{\uptau}$ is shown in parentheses. The last column shows the analytically derived coefficients for the NGM (see Section~\ref{sec:gm}).} \label{table:2d-fhit tau coeff}
\begin{tabular}{|c|c|c|c|c|c|c|c|c|c|}
\hline
\multirow{2}{*}{Case} & 
\multirow{2}{*}{Filter} & 
\multicolumn{2}{c|}{$\tau_{xx}$} & 
\multicolumn{2}{c|}{$\tau_{xy}$} & 
\multicolumn{2}{c|}{$\tau_{yy}$} & 
\multirow{2}{*}{Mean} & \multirow{2}{*}{NGM} \\ \cline{3-8}
\rule{0pt}{5ex}
& & $\left(\frac{\partial \filtered{u}}{\partial x}\right)^2 $ &  $\left(\frac{\partial \filtered{u}}{\partial y}\right)^2 $ &  $\frac{\partial \filtered{u}}{\partial x}\frac{\partial \filtered{v}}{\partial x}$ & $\frac{\partial \filtered{u}}{\partial y}\frac{\partial \filtered{v}}{\partial y}$ & $\left(\frac{\partial \filtered{v}}{\partial x}\right)^2 $ & $\left(\frac{\partial \filtered{v}}{\partial y}\right)^2 $ &  &  \\
\hline \hline

\multirow{6}{*}{K1}
& \multirow{2}{*}{Gaussian}
& \GMcoeff{11.53}{0.36} & \GMcoeff{11.83}{0.21} & \GMcoeff{11.72}{0.18} & \GMcoeff{11.8}{0.15} 
& \GMcoeff{11.84}{0.16} & \GMcoeff{11.63}{0.34} & \KJ{\GMcoeff{11.72}{0.27}} & {$\frac{\Delta^2}{12}$} \\
& & \multicolumn{2}{c|}{$\left(0.99\right)$} & \multicolumn{2}{c|}{$\left(0.99\right)$} & \multicolumn{2}{c|}{$\left(0.99\right)$} & 
$\left(0.99\right)$ & \\ 
\cline{3-10}

\rule{0pt}{4ex}
& \multirow{2}{*}{Box}
& \GMcoeff{11.3}{0.43} & \GMcoeff{11.42}{0.54} & \GMcoeff{11.35}{0.43} & \GMcoeff{11.43}{0.43} 
& \GMcoeff{11.4}{0.48} & \GMcoeff{11.38}{0.43} & \KJ{\GMcoeff{11.38}{0.46}} &  {$\frac{\Delta^2}{12}$}\\
& & \multicolumn{2}{c|}{$\left(0.99)\right)$} & \multicolumn{2}{c|}{$\left(0.99\right)$} & \multicolumn{2}{c|}{$\left(1.00\right)$} & 
$\left(0.99\right)$ &  \\
\cline{3-10}

\rule{0pt}{4ex}
& \multirow{2}{*}{Gaussian + box}
& \GMcoeff{5.63}{0.28} & \GMcoeff{5.76}{0.26} & \GMcoeff{5.79}{0.17} & \GMcoeff{5.78}{0.16} 
& \GMcoeff{5.79}{0.19} & \GMcoeff{5.67}{0.28} & \KJ{\GMcoeff{5.73}{0.24}} &  {$\frac{\Delta^2}{6}$}\\
& & \multicolumn{2}{c|}{$\left(0.99\right)$} & \multicolumn{2}{c|}{$\left(0.99\right)$} & \multicolumn{2}{c|}{$\left(0.99\right)$} & 
$\left(0.99\right)$ &  \\
\hline \hline

\multirow{6}{*}{K2}
& \multirow{2}{*}{Gaussian}
& \GMcoeff{10.87}{0.44} & \GMcoeff{11.33}{0.37} & \GMcoeff{11.58}{0.15} & \GMcoeff{11.58}{0.15} 
& \GMcoeff{11.33}{0.38} & \GMcoeff{10.88}{0.44} & \KJ{\GMcoeff{11.26}{0.45}} & $\frac{\Delta^2}{12}$\\
& & \multicolumn{2}{c|}{$\left(0.99\right)$} & \multicolumn{2}{c|}{$\left(0.98\right)$} & \multicolumn{2}{c|}{$\left(0.99\right)$} & 
$\left(0.99\right)$ &  \\
\cline{3-10}

\rule{0pt}{4ex}
& \multirow{2}{*}{Box}
& \GMcoeff{10.87}{0.46} & \GMcoeff{11.49}{0.39} & \GMcoeff{10.71}{0.46} & \GMcoeff{10.72}{0.46}
& \GMcoeff{11.5}{0.38} & \GMcoeff{10.64}{0.45} & \KJ{\GMcoeff{10.95}{0.58}} & $\frac{\Delta^2}{12}$\\
& & \multicolumn{2}{c|}{$\left(0.99\right)$} & \multicolumn{2}{c|}{$\left(0.98\right)$} & \multicolumn{2}{c|}{$\left(0.99\right)$} & 
$\left(0.99\right)$ & \\
\cline{3-10}

\rule{0pt}{4ex}
& \multirow{2}{*}{Gaussian + box}
& \GMcoeff{5.44}{0.31} & \GMcoeff{5.37}{0.24} & \GMcoeff{5.56}{0.15} & \GMcoeff{5.56}{0.15} 
& \GMcoeff{5.38}{0.24} & \GMcoeff{5.44}{0.22} & \KJ{\GMcoeff{5.46}{0.18}} & $\frac{\Delta^2}{6}$\\
& & \multicolumn{2}{c|}{$\left(0.99\right)$} & \multicolumn{2}{c|}{$\left(0.97\right)$} & \multicolumn{2}{c|}{$\left(0.98\right)$} & 
$\left(0.98\right)$ & \\
\hline \hline

\multirow{6}{*}{K3}
& \multirow{2}{*}{Gaussian}
& \GMcoeff{11.5}{0.37} & \GMcoeff{11.81}{0.22} & \GMcoeff{11.7}{0.18} & \GMcoeff{11.78}{0.14} 
& \GMcoeff{11.85}{0.14} & \GMcoeff{11.66}{0.30} & \KJ{\GMcoeff{11.72}{0.27}} & $\frac{\Delta^2}{12}$\\
& & \multicolumn{2}{c|}{$\left(0.98\right)$} & \multicolumn{2}{c|}{$\left(0.99\right)$} & \multicolumn{2}{c|}{$\left(0.99\right)$} & 
$\left(0.99\right)$ & \\
\cline{3-10}

\rule{0pt}{4ex}
& \multirow{2}{*}{Box}
& \GMcoeff{11.39}{0.53} & \GMcoeff{11.75}{0.53} & \GMcoeff{11.33}{0.31} & \GMcoeff{11.4}{0.41}
& \GMcoeff{11.40}{0.46} & \GMcoeff{11.38}{0.41} & \KJ{\GMcoeff{11.37}{0.45}} & $\frac{\Delta^2}{12}$\\
& & \multicolumn{2}{c|}{$\left(0.99\right)$} & \multicolumn{2}{c|}{$\left(0.99\right)$} & \multicolumn{2}{c|}{$\left(0.99\right)$} & 
$\left(0.99\right)$ & \\
\cline{3-10}

\rule{0pt}{4ex}
& \multirow{2}{*}{Gaussian + box}
& \GMcoeff{5.61}{0.28} & \GMcoeff{5.75}{0.25} & \GMcoeff{5.73}{0.18} & \GMcoeff{5.78}{0.16}
& \GMcoeff{5.79}{0.19} & \GMcoeff{5.70}{0.24} & \KJ{\GMcoeff{5.73}{0.23}} & $\frac{\Delta^2}{6}$\\
& & \multicolumn{2}{c|}{$\left(0.99\right)$} & \multicolumn{2}{c|}{$\left(0.99\right)$} & \multicolumn{2}{c|}{$\left(0.99\right)$} & 
$\left(0.99\right)$ & \\
\hline
\end{tabular}
\end{table}
\end{landscape}

\begin{table}[tp]
\centering
\caption {
Coefficients in Eq.~\eqref{eq:rbc tau discovered model}, the robustly discovered closure for $\bm{\uptau}$ for RBC (note that $\Delta^2$ is included in the coefficients). For Cases R1-R3 and different filter types, the mean and standard deviation of the discovered coefficients over different $N_{\text{LES}}$ are reported (see Table~\ref{table:rbc cases}). The average CC of the closure for each element of $\bm{\uptau}$ is shown in parentheses. The last column shows the analytically derived coefficients for the NGM (see Section~\ref{sec:gm}).} 
\label{table:rbc tau coeff}
\begin{tabular}{|c|c|c|c|c|c|c|} 
\hline
\multirow{2}{*}{Case} & 
\multirow{2}{*}{Filter} & $\tau_{xx}$ & $\tau_{xy}$ & $\tau_{yy}$ & 
\multirow{2}{*}{Mean} & \multirow{2}{*}{NGM} \\ \cline{3-5}
\rule{0pt}{5ex}
& & $\left(\frac{\partial \filtered{u}}{\partial x} \right)^2$ &  $\frac{\partial \filtered{u}}{\partial x} \frac{\partial \filtered{w}}{\partial x} $ & $\left(\frac{\partial \filtered{w}}{\partial x} \right)^2$  & & \\
\hline \hline

\multirow{6}{*}{R1}
& \multirow{2}{*}{Gaussian}
& \GMcoeff{10.89}{0.39} & \GMcoeff{11.01}{0.41} & \GMcoeff{10.55}{0.75} & \GMcoeff{10.98}{0.49}  & {$\frac{\Delta^2}{12}$} \\
& & $\left(0.98\right)$ & $\left(0.97\right)$ & $\left(0.92\right)$ & 
$\left(0.95\right)$ & \\ \cline{3-7}

\rule{0pt}{4ex}
& \multirow{2}{*}{Box}
& \GMcoeff{10.45}{0.94} & \GMcoeff{10.21}{0.93} & \GMcoeff{10.32}{0.87} & \GMcoeff{10.35}{0.97}  & {$\frac{\Delta^2}{12}$} \\
& & $\left(0.98\right)$ & $\left(0.94\right)$ & $\left(0.91\right)$ & 
$\left(0.93\right)$ & \\ \cline{3-7}

\rule{0pt}{4ex}
& \multirow{2}{*}{Gaussian + box}
& \GMcoeff{5.35}{0.56} & \GMcoeff{5.37}{0.35} & \GMcoeff{5.21}{0.48} & \GMcoeff{5.29}{0.66}  & {$\frac{\Delta^2}{6}$} \\
& & $\left(0.93\right)$ & $\left(0.94\right)$ & $\left(0.89\right)$ & 
$\left(0.91\right)$ & \\ 
\hline \hline

\multirow{6}{*}{R2}
& \multirow{2}{*}{Gaussian}
& \GMcoeff{11.35}{0.41} & \GMcoeff{11.82}{0.36} & \GMcoeff{9.7}{0.54} & \GMcoeff{10.62}{0.79}  & {$\frac{\Delta^2}{12}$} \\
& & $\left(0.98\right)$ & $\left(0.88\right)$ & $\left(0.81\right)$ & 
$\left(0.89\right)$ & \\ \cline{3-7}

\rule{0pt}{4ex}
& \multirow{2}{*}{Box}
& \GMcoeff{10.52}{0.65} & \GMcoeff{9.38}{0.5} & \GMcoeff{9.11}{0.59} & \GMcoeff{10.01}{0.44}  & {$\frac{\Delta^2}{12}$} \\
& & $\left(0.97\right)$ & $\left(0.90\right)$ & $\left(0.86\right)$ & 
$\left(0.91\right)$ & \\ \cline{3-7}

\rule{0pt}{4ex}
& \multirow{2}{*}{Gaussian + box}
& \GMcoeff{5.48}{0.24} & \GMcoeff{5.33}{0.12} & \GMcoeff{5.00}{0.23} & \GMcoeff{5.27}{0.28}  & {$\frac{\Delta^2}{6}$} \\
& & $\left(0.98\right)$ & $\left(0.92\right)$ & $\left(0.93\right)$ & 
$\left(0.94\right)$ & \\
\hline \hline

\multirow{6}{*}{R3}
& \multirow{2}{*}{Gaussian}
& \GMcoeff{11.22}{0.16} & \GMcoeff{11.34}{0.41} & \GMcoeff{10.51}{1.03} & \GMcoeff{11.02}{0.79}  & {$\frac{\Delta^2}{12}$} \\
& & $\left(0.94\right)$ & $\left(0.93\right)$ & $\left(0.91\right)$ & 
$\left(0.93\right)$ & \\ \cline{3-7}

\rule{0pt}{4ex}
& \multirow{2}{*}{Box}
& \GMcoeff{10.17}{0.32} & \GMcoeff{9.94}{0.64} & \GMcoeff{9.44}{1.32} & \GMcoeff{9.85}{0.95}  & {$\frac{\Delta^2}{12}$} \\
& & $\left(0.93\right)$ & $\left(0.93\right)$ & $\left(0.92\right)$ & 
$\left(0.92\right)$ & \\ \cline{3-7}

\rule{0pt}{4ex}
& \multirow{2}{*}{Gaussian + box}
& \GMcoeff{5.46}{0.10} & \GMcoeff{5.55}{0.12} & \GMcoeff{4.87}{0.66} & \GMcoeff{5.3}{0.54}  & {$\frac{\Delta^2}{6}$} \\
& & $\left(0.93\right)$ & $\left(0.90\right)$ & $\left(0.88\right)$ & 
$\left(0.90\right)$ & \\\hline 
\end{tabular}
\end{table}

\begin{table}[tp]
\centering
\caption{Coefficients in Eq.~\eqref{eq:rbc J discovered model}, the robustly discovered closure for $\bm{J}$ for RBC (note that $\Delta^2$ is included in the coefficients). For Cases R1-R3 and different filter types, the mean and standard deviation of the discovered coefficients over different $N_{\text{LES}}$ are reported (see Table~\ref{table:rbc cases}). The average CC of the closure for each element of $\bm{J}$ is shown in parentheses. The last column shows the analytically derived coefficients for the NGM (see Section~\ref{sec:gm})} \label{table:rbc j coeff}
\begin{tabular}{|c|c|c|c|c|c|} 
\hline
\multirow{2}{*}{Case} & 
\multirow{2}{*}{Filter} & $J_{x}$ & $J_{z}$ &  
\multirow{2}{*}{Mean} & \multirow{2}{*}{NGM} \\ \cline{3-4}
\rule{0pt}{4ex}  
& & $\frac{\partial \filtered{u}}{\partial x}\frac{\partial \filtered{\theta}}{\partial x} $ &  $\frac{\partial \filtered{w}}{\partial x}\frac{\partial \filtered{\theta}}{\partial x} $ &  & \\
\hline \hline

\multirow{6}{*}{R1}
& \multirow{2}{*}{Gaussian}
& \GMcoeff{10.54}{0.66} & \GMcoeff{10.3}{0.87} & \GMcoeff{10.88}{1.3}  & {$\frac{\Delta^2}{12}$} \\
& & $\left(0.93\right)$ & $\left(0.90\right)$ &  $\left(0.92\right)$ & \\ \cline{3-6}

\rule{0pt}{4ex}
& \multirow{2}{*}{Box}
& \GMcoeff{9.11}{0.86} & \GMcoeff{9.00}{0.65} & \GMcoeff{9.05}{0.80}  & {$\frac{\Delta^2}{12}$} \\
& & $\left(0.93\right)$ & $\left(0.92\right)$ &  $\left(0.93\right)$ & \\ \cline{3-6}

\rule{0pt}{4ex}
& \multirow{2}{*}{Gaussian + box}
& \GMcoeff{5.32}{0.3} & \GMcoeff{5.31}{0.5} & \GMcoeff{5.31}{0.45}  & {$\frac{\Delta^2}{6}$} \\
& & $\left(0.96\right)$ & $\left(0.90\right)$ & $\left(0.93\right)$ & \\ 
\hline \hline

\multirow{6}{*}{R2}
& \multirow{2}{*}{Gaussian}
& \GMcoeff{11.27}{0.2} & \GMcoeff{10.9}{0.4} & \GMcoeff{11.12}{0.37}  & {$\frac{\Delta^2}{12}$} \\
& & $\left(0.89\right)$ & $\left(0.85\right)$ & $\left(0.87\right)$ & \\ \cline{3-6}

\rule{0pt}{4ex}
& \multirow{2}{*}{Box}
& \GMcoeff{9.7}{0.11} & \GMcoeff{9.3}{0.23} & \GMcoeff{9.5}{0.67}  & {$\frac{\Delta^2}{12}$} \\
& & $\left(0.90\right)$ & $\left(0.84\right)$ &  $\left(0.87\right)$& \\ \cline{3-6}

\rule{0pt}{4ex}
& \multirow{2}{*}{Gaussian + box}
& \GMcoeff{5.55}{0.08} & \GMcoeff{5.1}{0.22} & \GMcoeff{5.32}{0.78}  & {$\frac{\Delta^2}{6}$} \\
& & $\left(0.91\right)$ & $\left(0.85\right)$ &  $\left(0.88\right)$ & \\ 
\hline \hline

\multirow{6}{*}{R3}
& \multirow{2}{*}{Gaussian}
& \GMcoeff{9.75}{0.47} & \GMcoeff{9.21}{0.34} & \GMcoeff{9.46}{0.97}  & {$\frac{\Delta^2}{12}$} \\
& & $\left(0.84\right)$ & $\left(0.83\right)$ &  $\left(0.83\right)$ & \\ \cline{3-6}

\rule{0pt}{4ex}
& \multirow{2}{*}{Box}
& \GMcoeff{9.87}{0.23} & \GMcoeff{9.5}{0.22} & \GMcoeff{9.68}{0.57}  & {$\frac{\Delta^2}{12}$} \\
& & $\left(0.80\right)$ & $\left( 0.81\right)$ &  $\left(0.81\right)$ & \\ \cline{3-6}

\rule{0pt}{4ex}
& \multirow{2}{*}{Gaussian + box}
& \GMcoeff{4.78}{0.12} & \GMcoeff{4.52}{0.34} & \GMcoeff{4.65}{0.77}  & {$\frac{\Delta^2}{6}$} \\
& & $\left(0.83\right)$ & $\left(0.80\right)$ &  $\left(0.81\right)$ & \\
\hline
\end{tabular}
\end{table}

\subsection{The Nonlinear Gradient Model (NGM): Taylor-series Expansion of the SGS Term} \label{sec:gm}
A closer examination of Eq.~\eqref{eq:2d-fhit tau discovered model} reveals that this closure is indeed the NGM (this includes both the form and the coefficients, within the uncertainty range). This connection was already pointed out by \citeA{zanna2020data}, although the implications and findings such as 1-4 mentioned in the previous subsection were not further discussed in their short letter.

First, let's briefly review the NGM \cite{leonard1975energy,clark1979evaluation, sagaut2006large}. As a simple illustration of the idea behind this model, \KJ{we have reproduced the derivation of NGM in \ref{appendix: GM derivation 1d} using a 1D arbitrary field, $a(x)$, for the reader's convenience}. Taylor-series expansion of $a(x-r_x)$ around $a(x)$ (Eq.~\eqref{eq:1D taylor}) simplifies the convolution integral of the filtering operation (Eq.~\eqref{eq:1dfilter}) such that $\overline{a}(x)$ can be written in terms of $a(x)$ and its derivatives, with coefficients that depend only on the moments of the filter's kernel, $G$ (Eq.~\eqref{eq:afiltered}). Using $\overline{u}^2$ and $\overline{u^2}$ as $a(x)$, we eventually arrive at an analytically derived closure for $\tau_{xx}$ with error $\mathcal{O}\left(\Delta^4\right)$ (Eq.~\eqref{eq:GM21D}). In 2D with filtering applied in both directions (like our 2D-FHIT), the NGM is \cite{sagaut2006large}   
\begin{eqnarray}
    {\bm{\uptau}}^\text{NGM}_{\rm{2D}} = \begin{bmatrix}
        \tau_{xx} & \tau_{xy} \\
        \tau_{yx} & \tau_{yy}
    \end{bmatrix}
    = c_{\tau}\Delta^2
    \begin{bmatrix}
        \left(\frac{\partial \filtered{u}}{\partial x}\right)^2 + \left(\frac{\partial \filtered{u}}{\partial y}\right)^2 &
        \frac{\partial \filtered{u}}{\partial x} \frac{\partial \filtered{v}}{\partial x} +
        \frac{\partial \filtered{u}}{\partial y} \frac{\partial \filtered{v}}{\partial y} \\ 
        \frac{\partial \filtered{u}}{\partial x} \frac{\partial \filtered{v}}{\partial x} +
        \frac{\partial \filtered{u}}{\partial y} \frac{\partial \filtered{v}}{\partial y}  &
        \left(\frac{\partial \filtered{v}}{\partial x}\right)^2 + \left(\frac{\partial \filtered{v}}{\partial y}\right)^2 
    \end{bmatrix}
    + \mathcal{O}\left(\Delta^4\right),
\label{eq:tau 2d-fhit taylor series expansion gradient model}
\end{eqnarray}
where $c_\tau$ depends on the filter's kernel. Similarly, for the 2D RBC with filtering only in the $x$ direction, the NGM is
\begin{eqnarray}
   {\bm{\uptau}}^\text{NGM}_{\rm{1D}} =  \begin{bmatrix}
        \tau_{xx} & \tau_{xz} \\
        \tau_{zx} & \tau_{zz}
    \end{bmatrix}
    = d_{\tau}\Delta^2
    \begin{bmatrix}
        \left(\frac{\partial \filtered{u}}{\partial x}\right)^2 &
        \frac{\partial \filtered{u}}{\partial x} \frac{\partial \filtered{w}}{\partial x} \\ 
        \frac{\partial \filtered{u}}{\partial x} \frac{\partial \filtered{w}}{\partial x} &
        \left(\frac{\partial \filtered{w}}{\partial x}\right)^2
    \end{bmatrix} 
    + \mathcal{O}\left(\Delta^4\right). \label{eq:tau rbc taylor series expansion gradient model}
\end{eqnarray}
As emphasized in \ref{appendix: GM derivation 1d}, there is nothing specific to momentum flux or even turbulence (or even physical systems) in the derivation of NGM. In fact, for the {\it filtered quadratic nonlinearity of any two arbitrary variables}, one arrives at the same expression with $\mathcal{O}\left(\Delta^4\right)$ accuracy. For example, following this derivation, for the SGS heat flux, we obtain
\begin{eqnarray}
    {\bm{J}}^\text{NGM}  = \begin{bmatrix} 
        J_x \\
        J_z
    \end{bmatrix}
    = d_{J}\Delta^2
    \begin{bmatrix}  
        \frac{\strut \partial \filtered{u}}{\strut \partial x} \frac{\strut \partial \filtered{\theta}}{\strut \partial x} \\
        \frac{\strut \partial \filtered{w}}{\strut \partial x} \frac{\strut \partial \filtered{\theta}}{\strut \partial x}
    \end{bmatrix}
    + \mathcal{O}\left(\Delta^4\right), \label{eq:J rbc taylor series expansion gradient model} 
\end{eqnarray}
where like $c_\tau$ and $d_{\tau}$, $d_J$ only depends on the filter's kernel. \KJ{Note that we are referring to NGM as a general class of closure models, representing the leading term of the Taylor-series expansion of the SGS term (whether it is for momentum or heat or any other flux).}

Computing $c_\tau$, $d_{\tau}$, $d_J$ for each of the filter types used in this study, we confirm that the discovered closures for the SGS stress are basically the NGM (Eqs.~\eqref{eq:tau 2d-fhit taylor series expansion gradient model}-\eqref{eq:tau rbc taylor series expansion gradient model}), and in the case of the SGS heat flux, an \KJ{NGM} (Eq.~\eqref{eq:J rbc taylor series expansion gradient model}) closure (see \crefrange{table:2d-fhit tau coeff}{table:rbc j coeff}). 

Based on the above analyses, we can now explain the findings (1)-(4) in Section~\ref{subsec:discovered model}. Closures of the same structure are robustly discovered for both SGS momentum and heat fluxes in two vastly different turbulent flows (and independent of parameters such as $Re$, $Ra$, $Pr$, and forcing) because the first term in the Taylor-series expansion dominates the SGS flux. As a result, in equation-discovery using common loss functions such as MSE and evaluation metrics such as CC, which aim at closely matching $\tau$ or $J$, NGM or NGM-like closures are discovered (if the library is expansive enough to include all the relevant terms). We emphasize that this would be the case with discovering the representation of the filtered nonlinearity of any two arbitrary variables. As already observed, the coefficients of the discovered closure become even closer to those of NGM as $\Delta$ is decreased (thus reducing potential contributions from the truncated $\mathcal{O}\left(\Delta^4\right)$ terms). 

The connection to the analytical derivation also explains why the coefficients in the discovered models are {\it independent} of the fluid or even the flow properties ($Ra$, $Re$, $Pr$) and only depend on the filter size ($\Delta$) and filter type. For the Gaussian and box filters we obtain $c_\tau = d_{\tau} = d_J = 1/12$: this is because the parameters of the Gaussian filter are chosen such that Gaussian and box filters' kernels have the same second moment \KJ{(see Eq. \eqref{eq:1D filter moments} for the definition of moment)} \cite{pope2000turbulent}. For the Gaussian + box filter, the coefficients are $1/6$ because the kernel of this filter is convolution of the Gaussian and box filter kernels. For the sharp-spectral filter, the moments are indefinite, this is why there is no NGM discovery with this filter (and we will discuss later why the equation discovery fails altogether). Note that coarse-graining done here via cutoff in the spectral space does not change $c_{\tau}, d_{\tau}$ and $d_{\tau}$; however, if coarse-graining is done by other techniques such as box averaging, then the coefficients might change (note that the NGM coefficient for Gaussian~+~box filter is \KJnew{twice} the coefficient of either filter, \KJnew{as the coefficients are additive}; see Tables~\ref{table:2d-fhit tau coeff}-\ref{table:rbc j coeff}). 

In short, one can explain the effects of different filter kernels and coarse-graining strategies on the discovered closures following the analytically derivable NGM (see \ref{appendix: GM derivation 1d} and \citeA{sagaut2006large}).

\KJ{Note that as mentioned earlier, reconstructing SGS fluxes from the filtered solution is a solvable problem as long as the filtering operation is reversible. This is the case for filters with compact support, which include Gaussian, box, and Gaussian + box filters, but not sharp-spectral (cutoff) filters. However, once coarse-graining is applied in addition to any of these filters, the full operation is no longer reversible. Still, the part of the spectrum above the Nyquist frequency can be recovered. It should be noted that NGM is an approximation of the deconvolution of SGS fluxes \cite{sagaut2006large}. The deconvolution procedure can be done only for invertible filters, i.e., non-projective filters. Projective filters, such as sharp-spectral cutoffs, induce an irreversible loss of information, rendering the original data unrecoverable. This is another way of understanding why the NGM exists for Gaussian and box filters (and their combinations) but not for sharp-spectral cutoff filters.}

An important implication of the above findings and discussions is that the discovered closure may not be unique and can depend on the filtering and coarse-graining procedure: it depends on  the filter type (and up to a factor, on the filter size). This is not a problem of equation-discovery; in fact, the SGS fluxes diagnosed from FDNS are not unique and depend on the filtering and coarse-graining procedure  (this is further shown in \cref{fig: snapshot 2dfhit different filters} and discussed at the end of this section). This has implications not just for equation-discovery, but more broadly, for the ongoing efforts on learning SGS parameterizations for various processes from high-fidelity data using ML. See \citeA{sun2023quantifying} for extensive discussions about this issue focused on the data-driven SGS modeling of atmospheric gravity waves.

The next key question is about the accuracy and stability of LES of the 2D-FHIT and RBC with the NGM closures, $\bm{\uptau}^\text{NGM}$ and $\bm{J}^\text{NGM}$. However, before discussing the \aposteriori (online) performance of NGM closures, we address one more issue, and that is about any potential influence from numerical calculations in our equation discovery.

\subsubsection{Effects of numerical discretization}
The appearance of gradients of velocity (or temperature) in Eqs.~\eqref{eq:tau 2d-fhit taylor series expansion gradient model}-\eqref{eq:J rbc taylor series expansion gradient model} might suggest to some that the discovered equations represent the truncated terms of finite difference/volume discretization schemes (the methods used in \citeA{zanna2020data}). The discussions in their paper and the comprehensive analyses here should leave no ambiguity that Eqs.~\eqref{eq:tau 2d-fhit taylor series expansion gradient model}-\eqref{eq:J rbc taylor series expansion gradient model} represent the physics of the SGS fluxes, rather than any numerical error. Still, we wish to discuss a few more points here, as numerical errors from finite difference/volume discretizations or from aliasing (in spectral calculations) can certainly contaminate equation discovery.  

All numerical calculations in this study are performed using Fourier and Chebyshev spectral methods. Moreover, we have repeated our calculations of the SGS fluxes and of the basis functions in the library after de-aliasing based on the $2/3$ rule \cite{orszag1971elimination}. Furthermore, we have \KJ{repeated the} discovery on fluxes that are only filtered but not coarse-grained (thus they remain on the high-resolution DNS grid). The outcomes of all these experiments are Eqs.~\eqref{eq:tau 2d-fhit taylor series expansion gradient model}-\eqref{eq:J rbc taylor series expansion gradient model}, demonstrating that the discovered closures do not contain any contributions from numerical errors.  

\subsection{{\bf {\it \textbf{A posteriori}}} (Online) Tests and Inter-scale Energy/Enstrophy Transfer}
For all 6 cases and all tested $N_\text{LES}$, the LES runs with NGM closures are unstable: High-wave number features appear and the simulations eventually blow up (not shown). This is consistent with the findings of \citeA{zanna2020data}, who only found stable LES once the SGS momentum fluxes predicted by the discovered closure were attenuated \KJ{(also see recent work by \citeA{perezhogin2023implementation} who showed that further filtering the output of NGM can lead to promising online results in an ocean model)}. More generally, this is also consistent with extensive studies in the 1990s (though mainly focused on 3D turbulence), which found that LES with the NGM closure is unstable \cite{liu1994properties, leonard1997large,vreman1997large, borue1998local, meneveau2000scale, pope2000turbulent, chen2003physical,chen2006physical}. The exact reason(s) for the instabilities remain unclear but these studies found that in general, in 3D turbulence, NGM has insufficient dissipation and/or too much backscattering; see, e.g., the discussions in \citeA{leonard1997large,leonard2016large} and \citeA{sagaut2006large}. \KJ{Over the years, several methods, including positive clipping (removing backscattering), regularization, modulation, and the dynamic procedure, have been employed to overcome this instability problem and enhance the NGM's performance in \textit{a posteriori} tests \cite{balarac2013dynamic, liu1994properties, prakash2022optimal, khani2017modulated, khani2022gradient, khanigradient}. Later} studies focused more on eddy-viscosity closures, or on NGM combined with eddy-viscosity, the so-called mixed models \cite{winckelmans1998testing, cottet1996artificial, balarac2013dynamic}. Such versions of NGM have been used in some geophysical flows, e.g., for atmospheric boundary layer~\cite{lu2010modulated, lu2014development, khani2017modulated, khani2020anisotropic, khani2022gradient} and oceanography \cite{khanigradient}.

In 2D turbulence with filtering done in both directions, such as our 2D-FHIT cases, \KJ{the NGM has a clear major shortcoming: it cannot capture any energy transfer between the subgrid and resolved scales, despite capturing the enstrophy transfer well} \cite{chen2003physical, chen2006physical, nadiga2008orientation}. To further explore this issue, first note that the rate of kinetic energy transfer between the resolved and subgrid scales, $P_{\tau}$, is \cite{pope2000turbulent} 
\begin{eqnarray} \label{eq: sgs P}
    P_{\tau} = -\tau_{ij}^r \filtered{S}_{ij},
\end{eqnarray}
where summation over repeated indices is implied. $\filtered{S}$ and ${\tau^r}$ are the 2D filtered rate of strain tensor and the anisotropic part of the SGS stress tensor (see \ref{appendix: sgs energy transfer} for details). \KJ{In homogeneous 2D turbulence, globally (domain averaged), there is no net forward cascade of kinetic energy, which has been used to develop successful parameterizations that are globally energetically consistent~\cite{boffetta2012two, jansen2015energy}. However, locally (at a given grid point), there can be both forward transfer and backscatter of energy, which can be important to capture by the SGS closures~\cite{thuburn2014cascades}. In 2D turbulence with filtering done in both directions, using ${\bm{\uptau}}^\text{NGM}$ in the above equation shows that  $P^{\rm{NGM}}_{\tau}(x,y,t)$ is identically zero \KJ{at every grid point} (see \ref{appendix: sgs energy transfer}). This is demonstrated numerically in \Cref{table:2d CC}, which also shows that NGM captures both forward transfer and backscatter of SGS {\it enstrophy} fairly well (CC$>0.95$). Therefore, despite the high CC of ${\bm{\uptau}}^\text{NGM}$ with ${\bm{\uptau}}^\text{FDNS}$, and even a fairly accurate inter-scale {enstrophy} transfer, NGM cannot capture any inter-scale {\it energy} transfer, indicating a major failure from a functional modeling perspective (note that in this context, ``inter-scale'' means between the {\it resolved} and {\it subgrid} scales)}. A physical/mathematical interpretation of this failure is that while NGM reproduces the structure of $\bm{\uptau}$ remarkably well, it does not at all capture the correlations between the $\bm{\uptau}$ and $\filtered{\textbf{S}}$ tensors, e.g., the angles between their principle directions \cite{leonard2016large}.  

This inability to represent any inter-scale energy transfer is likely the reason for the instabilities of LES with NGM closure in Cases K1-K3 (and generally, in 2D turbulence). But how about in RBC? In Cases R1-R3, filtering is conducted only in the horizontal direction, and as a result, $P^{\rm{NGM}}_{\tau}$ is not identically zero. In fact, in these cases, the forward transfer and backscatter of both kinetic energy and enstrophy are captured fairly well by NGM, with CC often above $0.95$ (\Cref{table:rbc CC}). However, a deeper examination shows that the backscatter (anti-diffusion) of inter-scale SGS potential energy, measured as $P_J$ (see \ref{appendix: sgs energy transfer}), may not be captured well, specially at low $N_{\rm{LES}}$ (\cref{table:rbc CC}). Poor representation of backscattering can certainly lead to instabilities, as for example shown by \citeA{guan2022stable} for 2D turbulence. 

\begin{table}[t]
\centering
\caption {The average correlation coefficient (CC) between inter-scale energy transfer ($P_\tau$) or enstrophy transfer ($P_Z$) patterns of the SGS momentum stresses from FDNS and from NGM closure \KJ{(Eq. \eqref{eq:2d-fhit tau discovered model})} for Cases K1-K3 and different $N_{\rm{LES}}$. The CC of $P_{\tau}$ for both forward transfer ($>0$) and backscatter ($<0$) of SGS energy is ``undefined'' since $P^{\text{NGM}}_{\tau}=0$ everywhere for 2D-FHIT (in general, $P^{\text{FDNS}}_{\tau} \ne 0$). On the contrary, the forward transfer and backscatter of SGS enstrophy are captured well by the NGM. The Gaussian filter is used in FDNS.} 
\begin{tabular}{ |c|c|c|c|c| }
\hline
Cases & $N_{\text{LES}}=32 $ & $N_{\text{LES}}=64 $ & $N_{\text{LES}}=128 $          & $N_{\text{LES}}=256 $         \\
\hline \hline
\multicolumn{5}{c}{CC for $P_{\tau} \left(P_{\tau}>0, P_{\tau}<0\right)$} \\
\hline
K1&\multicolumn{4}{c|}{} \\
K2&\multicolumn{4}{c|}{undefined (undefined, undefined)} \\
K3&\multicolumn{4}{c|}{} \\
\hline
\multicolumn{5}{c}{CC for $P_{Z} \left(P_{Z}>0, P_{Z}<0\right)$} \\
\hline
K1 & $0.98 \left(0.98, 0.97 \right)$ & $0.98 \left(0.98, 0.97 \right)$ & $0.98 \left(0.98, 0.97 \right)$ & $0.98 \left(0.98, 0.96 \right)$\\
K2 & - & - & $0.98 \left(0.98, 0.97\right)$ & $0.99 \left(0.99, 0.98\right)$\\ 
K3 & $0.98 \left(0.98, 0.97\right)$ & $0.98 \left(0.98, 0.96\right)$ & $0.97 \left(0.97, 0.95\right)$ & $0.96 \left(0.97, 0.93\right)$\\
\hline
\end{tabular}
\label{table:2d CC}
\end{table}

\begin{table}[t]
\centering
\caption {The average correlation coefficient (CC) between inter-scale kinetic energy transfer ($P_\tau$) or enstrophy transfer ($P_Z$) or potential energy transfer ($P_J$) patterns of the SGS fluxes from FDNS and from NGM closure \KJ{(Eqs.~\eqref{eq:rbc tau discovered model}-\eqref{eq:rbc J discovered model})} for Cases R1-R3 and different $N_{\rm{LES}}$. Note that for RBC, filtering is conducted in only one direction ($x$), therefore, $P_{\tau}$ is not identically zero. Here, the forward transfer and backscatter of SGS kinetic energy and enstrophy are overall captured well, specially as $N_{\rm{LES}}$ increases. However, the backscatter of SGS potential energy is not well captured, specially at low LES resolutions. The Gaussian filter is used in FDNS. See \ref{appendix: sgs energy transfer} for the definition of $P_J$.} 
\begin{tabular}{ |c|c|c|c|c| } 
\hline
Cases & $N_{\text{LES}}=128 $ & $N_{\text{LES}}=256 $  & $N_{\text{LES}}=512 $        \\
\hline \hline
\multicolumn{4}{c}{CC for $P_{\tau} \left(P_{\tau}>0, P_{\tau}<0\right)$} \\
\hline
R1 & $0.94 \left(0.96, 0.85\right)$ &  $0.99 \left(0.99, 0.98 \right)$ & $-$\\
R2 & $0.97 \left(0.81, 0.98 \right)$& $0.98 \left(0.91, 0.98 \right)$ & $0.99 \left(0.97,1.00\right)$\\
R3 & $0.79 \left(0.81, 0.74\right)$ & $0.88 \left(0.92, 0.81\right)$ & $ 0.96 \left(0.97, 0.93 \right) $\\
\hline
\multicolumn{4}{c}{CC for $P_{Z} \left(P_Z>0, P_Z<0\right)$} \\
\hline
R1 & $1.00 \left(1.00, 0.99\right)$ &  $1.00 \left(1.00, 1.00 \right)$ & $-$\\
R2 & $0.99 \left(0.96, 1.00 \right)$& $0.99 \left(0.98, 1.00 \right)$ & $1.00 \left(1.00,1.00\right)$\\
R3 & $0.96 \left(0.95, 0.96\right)$ & $0.99 \left(0.99, 0.98\right)$ & $ 1.00 \left(1.00, 0.99 \right) $\\
\hline
\multicolumn{4}{c}{CC for $P_{J} \left(P_{J}>0, P_{J}<0\right)$} \\
\hline
R1 & $0.89 \left(0.89, 0.15\right) $ &  $0.97 (0.97, 0.46) $ & $-$\\
R2 & $0.76 \left(0.75, 0.65\right)$& $0.91 \left(0.91, 0.63\right)$ & $0.98 \left(0.98, 0.76\right)$\\
R3 & $0.77 \left(0.75, 0.40\right) $ & $ 0.87 \left(0.86, 0.39\right) $ & $ 0.94 \left(0.94, 0.44\right) $\\
\hline
\end{tabular}
\label{table:rbc CC}
\end{table}

\begin{figure}[t]
  \centering
  \includegraphics[width=\textwidth]{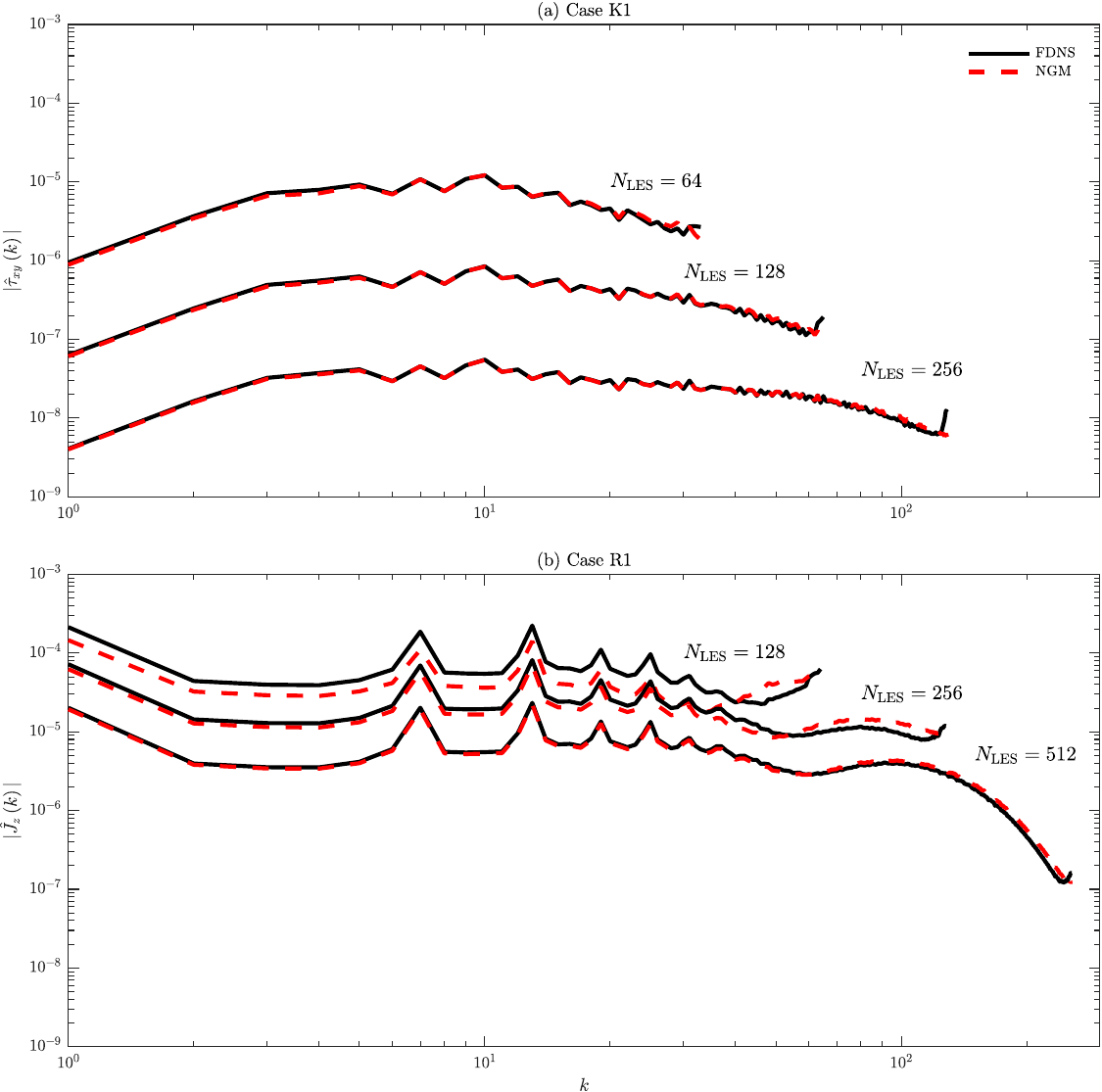}
  \caption{Examples of the spectra of SGS fluxes predicted using NGM compared to those diagnosed using FDNS (the truth). (a) $\tau_{xy}$ from Case K1 and (b) $J_z$ from Case R1 for 3 different $N_{\rm{LES}}$. A Gaussian filter is used for FDNS, but the same behavior is observed for box and Gaussian+box filters. Here, $|\hat{\cdot}|$ is the modulus of Fourier coefficients.}
  \label{fig: spectra GM SGS}
\end{figure}

To further explore other potential shortcomings of NGM, we have also examined the spectra of elements of $\bm{\uptau}^{\rm{NGM}}$ and $\bm{J}^{\rm{NGM}}$ in comparison to those from FDNS (\Cref{fig: spectra GM SGS}). This analysis shows that the spectra of SGS momentum and heat fluxes are captured well across wavenumbers, even at high wavenumbers, indicating that NGM performs well in this {\it a priori} (offline) metric.    

To summarize the above analyses: we find all LES with NGM closures for 2D-FHIT and RBC cases to become unstable even at high LES resolutions. Understanding the reason(s) for this poor {\it a posteriori} (online) performance is essential to make further progress. Examining a few functional and structural metrics beyond CC of SGS fluxes (e.g., inter-scale energy/enstrophy transfers, spectra) point to only one major shortcoming that is relevant to 2D-FHIT (and any 2D turbulent flow): NGM cannot capture any inter-scale kinetic energy transfer, which is likely the reason for the instabilities. This is not an issue in RBC, for which we only identify one shortcoming, and that is the poor representation of backscatter (anti-diffusion) of potential energy, specially at low LES resolution. These findings indicate that the poor {\it a posteriori} (online) performance of NGM might have different causes in different flows and requires more extensive investigations.

Before discussing ideas for addressing these challenges in future work, we present more analyses in two areas: a closer examination of the physics included in the library (Subsection~\ref{subsec: library}) and the decomposition of the SGS fluxes and the sensitivity of the diagnosed fluxes to the filter type/size (Subsection~\ref{subsubsec: taylor series expansion of leonard, cross and reynolds stresses}).

\subsection{A Physics-guided Library: Pope Tensors} \label{subsec: library}

In \cref{subsec:discovered model}, we consider an expansive library of basis functions combining the low- and high-order derivatives and polynomials of velocity and temperature. Under certain assumptions, smaller but physics-informed libraries can be devised. For example, \citeA{boussinesq1877essai} hypothesized that for a nearly homogeneous, incompressible, high-$Re$ flow, the anisotropic SGS stress $\bm{\uptau}^{\text{r}}$ (Eq.~\eqref{eq:taur}) is only a function of the filtered rates of strain $\filtered{\textbf{S}}$ (Eq.~\eqref{eq: filtered S}) and rotation $\filtered{\bm{\Upomega}}$ (Eq.~\eqref{eq:omega}) tensors:
\begin{eqnarray}
    \bm{\uptau}^{\text{r}} & = &\bm{\uptau}^{\text{r}}\left(\filtered{\textbf{S}}, \filtered{\bm{\Upomega}}\right), \label{eq:anisotropic stress}\\ 
    \filtered{\bm{\Upomega}} & = & \frac{1}{2}
    \begin{bmatrix}
    0 & \frac{\partial \filtered{u}}{\partial y} - \frac{\partial \filtered{v}}{\partial x} \\
    \frac{\partial \filtered{v}}{\partial x} - \frac{\partial \filtered{u}}{\partial y} & 0
    \end{bmatrix}. 
    \label{eq:omega}
\end{eqnarray}
Note that in Eqs.~\eqref{eq:2d-fhit filtered uv momentum} and \eqref{eq:rbc filtered uv momentum} and in general, only $\nabla \cdot \bm{\uptau}^{\text{r}}$ has to be parameterized as the rest of $\nabla \cdot \bm{\uptau}$ can be absorbed into $\nabla \filtered{p}$ \cite{sagaut2006large}. Owing to Cayley-Hamilton theorem \cite{gantmakher2000theory}, $\bm{\uptau}^{\text{r}}$ can be represented as a linear combination of a finite number of tensors, the so-called Pope tensors \cite{pope1975more}. In 2D, there are only 3 Pope tensors $\textbf{Z}$, thus
\begin{eqnarray}
    \bm{\uptau}^{\text{r}} = \sum_{n=0}^{2} \zeta^{\left(n\right)} \left(I_1,I_2\right)\textbf{Z}^{\left(n\right)}.
\end{eqnarray}
The three Pope's tensors are $\textbf{Z}^{\left(0\right)}=\textbf{I}$, $\textbf{Z}^{\left(1\right)}=\filtered{\textbf{S}}$, and
\begin{eqnarray}
    \textbf{Z}^{\left(2\right)} =\filtered{\textbf{S}}\,\boldsymbol{\filtered{\Omega}}  -\boldsymbol{\filtered{\Omega}}\,\filtered{\textbf{S}}
    = -\frac{1}{2}
    \begin{bmatrix}
        \left(\frac{\partial \filtered{u}}{\partial y}\right)^2 - \left(\frac{\partial \filtered{v}}{\partial x}\right)^2  &
        2\left(\frac{\partial \filtered{u}}{\partial x}\frac{\partial \filtered{v}}{\partial x} + \frac{\partial \filtered{u}}{\partial y}\frac{\partial \filtered{v}}{\partial y}\right)  \\ 
        2\left(\frac{\partial \filtered{u}}{\partial x}\frac{\partial \filtered{v}}{\partial x} + \frac{\partial \filtered{u}}{\partial y}\frac{\partial \filtered{v}}{\partial y}\right) &
        -\left(\frac{\partial \filtered{u}}{\partial y}\right)^2 + \left(\frac{\partial \filtered{v}}{\partial x}\right)^2
    \end{bmatrix},
\end{eqnarray}
which is related to the anisotropic part of the NGM stress. In fact, \KJ{$\bm{\uptau}^{\rm{NGM}^{\text{r}}}=-\Delta^2 \textbf{Z}^{\left(2\right)}/12$} (see Eq.~\eqref{eq:tauGMr}). Note that this is also the physics-based closure derived in \citeA{anstey2017deformation}. Coefficients $\zeta^{\left(n\right)}$ are functions of invariants $I_{1} = \text{tr}( \filtered{\textbf{S}}^{\,2})$ and $I_{2} = \text{tr}( \filtered{\bm{\Omega}}^{\,2})$. The standard Smagorinsky model is $\zeta^{(1)}(I_1) \textbf{Z}^1$.

Our expansive library, described in Eqs.~\eqref{eq:lib1}-\eqref{eq:lib2}, includes the individual terms to discover \KJ{$\textbf{Z}^{(n)}$ ($n=0,1,2$)}; however, we have always found the NGM stress, $\bm{\uptau}^{\rm{NGM}}$. To see whether the results will change with a discovery only done on the anisotropic SGS stress, $\bm{\uptau}^{\rm{r}}$, and with a smaller library that only has the terms relevant to the Pope tensors, we have conducted more experiments with 3 libraries for Cases K1-K3. The first library only includes the 3 Pope tensors \KJ{$\{\textbf{Z}^{(0)}, \textbf{Z}^{(1)}, \textbf{Z}^{(2)}\}$}, the second library only includes the 6 non-zero elements of these tensors, and the third library only includes the 8 terms that compromise these 6:
\begin{eqnarray}
    \Biggl\{1, \frac{\partial \filtered{u}}{\partial x},  \frac{\partial \filtered{u}}{\partial y}, \frac{\partial \filtered{v}}{\partial x}, \left(\frac{\partial \filtered{u}}{\partial y}\right)^2,  \left(\frac{\partial \filtered{v}}{\partial x}\right)^2, \frac{\partial \filtered{u}}{\partial x}\frac{\partial \filtered{v}}{\partial x}, \frac{\partial \filtered{u}}{\partial y}\frac{\partial \filtered{v}}{\partial y} \Biggr\}.
\end{eqnarray}
The RVM with any of these libraries robustly discovers \KJ{$\bm{\uptau}^{\rm{NGM^{\text{r}}}}=-\Delta^2 \textbf{Z}^{(2)}/12$}, without $\textbf{Z}^{(1)}$ (or $\textbf{Z}^{(0)}$) showing up (thus, no Smagorinsky/eddy viscosity-like term). Needless to say, LES with this closure is unstable. 

The above analyses show the prevalence of NGM: it emerges whether the full or just the anisotropic part of the SGS stress tensor is discovered, and whether an expansive or a small physics-guided library is used.

\subsection{Decomposition of SGS Fluxes: Leonard, Cross, and Reynolds Stresses}\label{subsubsec: taylor series expansion of leonard, cross and reynolds stresses}

As discussed earlier, whether a closure could be successfully discovered from FDNS data or if the NGM could be derived depend on the choice of the filter. The latter was explained based on the dependence of the derived closure on the moments' of the filter kernel. Furthermore, the coefficients of the discovered closure and the analytically derived coefficients of the NGM depend on the choice of the filters (Tables~\ref{table:2d-fhit tau coeff}-\ref{table:rbc j coeff}). Here, we further demonstrate the sensitivity of the diagnosed FDNS SGS flux (which is treated as truth in offline/supervised learning data-driven modeling approaches) to the choice of the filter, and then decompose the flux into its three components to gain further insight.    

\begin{figure}
  \includegraphics[width=\textwidth]{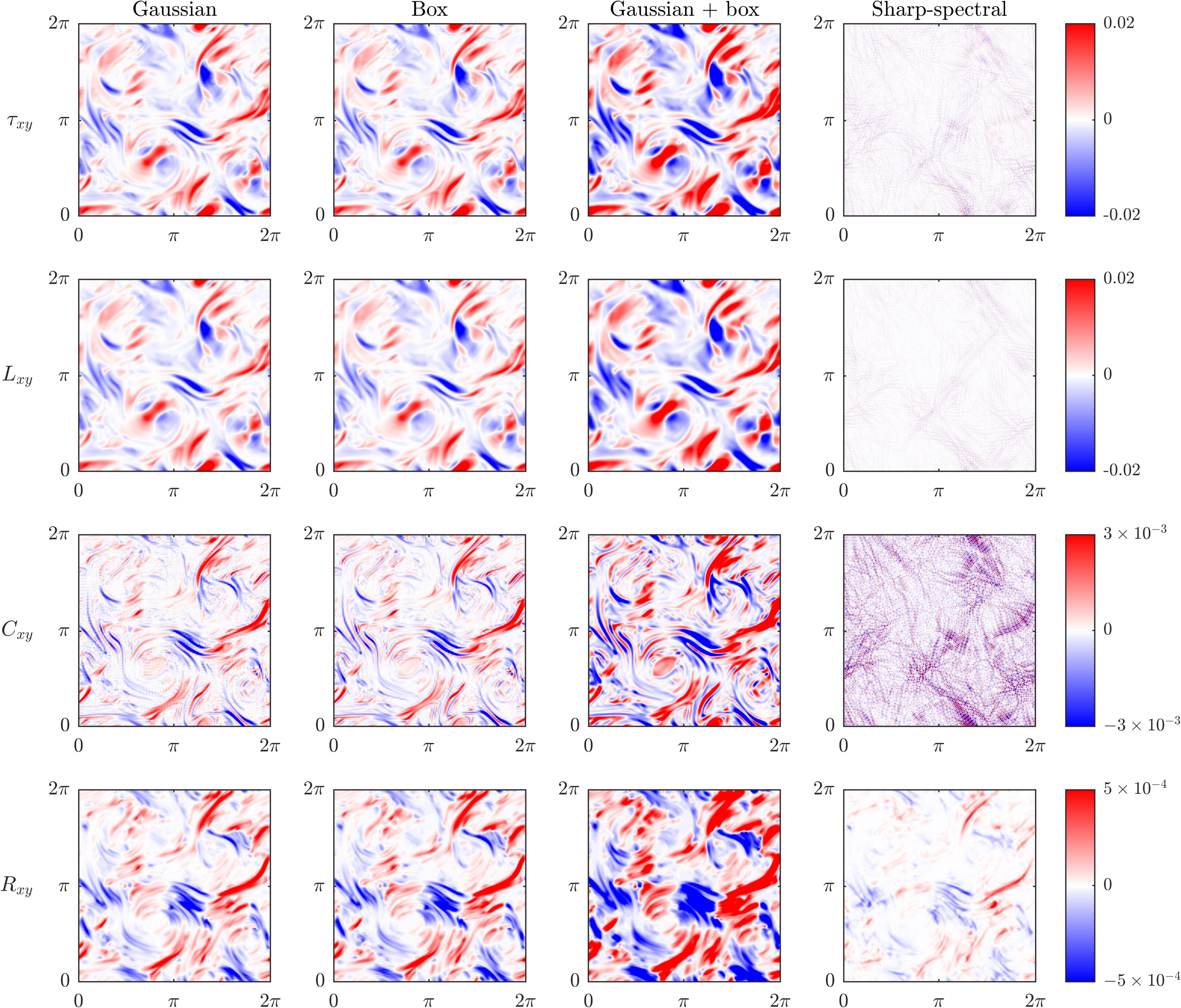}
  \justifying
  \caption{The first row shows examples of snapshots of the SGS stress, $\tau_{xy}$, for Case K1, diagnosed from FNDS data using different filters and $N_{\text{LES}}=128$ (see \cref{table:2d-fhit cases}). Rows 2-4 show the three components of this $\tau_{xy}$: the Leonard stress, $L_{xy}$, cross stress, $C_{xy}$, and Reynolds stress, $R_{xy}$. Note the substantially different ranges of the colorbars.}
  \label{fig: snapshot 2dfhit different filters}
\end{figure}

The top row of \cref{fig: snapshot 2dfhit different filters} shows examples of SGS $\uptau$ in 2D-FHIT diagnosed from the FDNS data using different filter types. It is clear that the diagnosed fluxes are not unique and particularly different between Gaussian/box filters and sharp-spectral filter (similar differences can be seen in SGS momentum and heat fluxes in RBC). This sensitivity, which has important implications for data-driven SGS modeling efforts \cite{sun2023quantifying}, has been known for a long time in the LES community \cite{leonard1975energy, sagaut2006large}. The Gaussian and box filters extract fairly similar features, even of almost the same amplitude (which is due to their matched kernels' second moments). The Gaussian+box filter captures similar features but with a factor of $\sim 2$ difference in amplitude (related to the factor of 2 difference in NGM coefficients). However, the sharp-spectral filter extracts very different features that have much smaller length scales and amplitudes. \KJ{We speculate that the failure of the discovery with the cutoff filter is a result of the inability of the current library in representing these features (at least in {\it sparsely} representing these features, if it exists).} We also point out that in \citeA{guan2022stable,guan2023learning},  deep convolutional neural networks (CNNs) could not be successfully trained on FDNS data obtained using sharp-spectral {\it cutoff} filters, while high CC and stable/accurate LES runs in different systems were achieved using CNNs trained on FDNS data obtained through the Gaussian filter. Note that \citeA{ross2023benchmarking} successfully trained CNNs (and performed equation-discovery) using a ``smoothed'' sharp-spectral filter that had exponential decay at high wavenumbers (rather than a cutoff). These findings further show the importance of how the ``true'' SGS fluxes are diagnosed for offline/supervised learning.

To see the reason for this difference, we decompose the SGS tensor using $\bm{u}=\overline{\bm{u}}+\bm{u}'$.  \citeA{leonard1975energy} introduced a decomposition of the SGS tensor into three components. However, since two of these components were not Galilean-invariant \cite{speziale1985galilean}, a Galilean-invariant decomposition was later proposed by \citeA{germano1986proposal}:
\begin{eqnarray}
    \bm{\uptau} = \textbf{L} + \textbf{C} + \textbf{R}.
\end{eqnarray}
Here, $\textbf{L}$, $\textbf{C}$, and $\textbf{R}$ are the Leonard, cross, and Reynolds stresses, which in 2D-FHIT are
\begin{eqnarray}
    \textbf{L} = 
    \begin{bmatrix}
    L_{xx} & L_{xy} \\
    L_{yx} & L_{yy}
    \end{bmatrix} =
    \begin{bmatrix}
    \,\,\filtered{\filtered{u}^{\,2}}-\filtered{\filtered{u}}^{\,2} &
    \filtered{\filtered{u}\,\filtered{v}}-\filtered{\filtered{u}}\,\filtered{\filtered{v}} \\ 
    \,\,\filtered{\filtered{u}\,\filtered{v}}-\filtered{\filtered{u}}\,\filtered{\filtered{v}}
    & \filtered{\filtered{v}^{\,2}}-\filtered{\filtered{v}}^{\,2}
    \end{bmatrix},
\end{eqnarray}
\begin{eqnarray}
    \textbf{C} = 
    \begin{bmatrix}
        C_{xx} & C_{xy} \\
        C_{yx} & C_{yy}
    \end{bmatrix} =
    \begin{bmatrix}
    \renewcommand*{\arraystretch}{1.5}
        2\left(\,\filtered{\filtered{u}u'} - \filtered{\filtered{u}}\,\filtered{u'}\,\right)  & 
        \filtered{\filtered{u}v'} + \filtered{u'\,\filtered{v}}- \filtered{\filtered{u}}\,\filtered{v'} - \filtered{u'}\,\filtered{\filtered{v}}\, \\
        \filtered{\filtered{u}v'} + \filtered{u'\,\filtered{v}}- \filtered{\filtered{u}}\,\filtered{v'} - \filtered{u'}\,\filtered{\filtered{v}}\, &
        2\left(\,\filtered{\filtered{v}v'} - \filtered{\filtered{v}}\,\filtered{v'}\,\right)
    \end{bmatrix},
\end{eqnarray}
\begin{eqnarray}
    \textbf{R} = 
    \begin{bmatrix}
        R_{xx} & R_{xy} \\
        R_{yx} & R_{yy}
    \end{bmatrix} =
    \begin{bmatrix}
        \,\,\filtered{u'^{\,2}} - \filtered{u'}^{\,2} & 
        \filtered{u'v'} - \filtered{u'}\,\filtered{v'}\\
        \filtered{u'v'} - \filtered{u'}\,\filtered{v'}
        & \filtered{v'^{\,2}} - \filtered{v'}^{\,2}
    \end{bmatrix}.
\end{eqnarray}
$\bm{\uptau}$ and $\bm{J}$ of RBC can be decomposed in the same fashion. The most familiar component, the Reynolds stress, represents interactions in the unresolved scales that project onto the resolved scale. The cross stress represents the direct interactions between the unresolved and resolved scales that project onto the resolved scale. The Leonard stress includes interactions between the resolved scales not captured by the low-resolution LES grid. See \citeA{leonard1975energy}, \citeA{mcdonough2007introductory}, \KJ{and  \citeA{perezhogin2023subgrid}} for more discussions.  

The relative importance of these three components in $\bm{\uptau}$ and $\bm{J}$ depends on the filter type and size (and even the flow characteristics). Rows 2-4 of \cref{fig: snapshot 2dfhit different filters} show examples of the Leonard, cross, and Reynolds stress components of $\tau_{xy}$. For Gaussian, box, and Gaussian+box filters, the Leonard stress dominates, followed by cross and then Reynolds stress. However, for sharp-spectral, only the Reynolds stress has coherent structures that more or less resemble the Reynolds stress from Gaussian/box filters. The strong dependence on filter type comes from the fact that for Gaussian and box filters, $\overline{\bm{u}'} \ne 0$ and $\overline{\overline{\bm{u}}} \ne \overline{\bm{u}}$, leading to non-zero Leonard and cross stresses. 

 However, for the sharp-spectral filters, $\overline{\bm{u}'} = 0$ and $\overline{\overline{\bm{u}}} = \overline{\bm{u}}$.  \KJnew{As a result, with de-aliasing applied, the (coarse-grained) Leonard stress may or may not be zero, depending on details of the calculations as discussed next (the cross stress is non-zero). This is because of the nuances in how the cutoff wavenumber for the sharp-spectral kernel is defined for both filtering and coarse-graining; for example, we find the coarse-grained Leonard stress to be non-zero in \cref{fig: snapshot 2dfhit different filters}}. Here, following the common approach in the literature \cite{pope2000turbulent}, for filtering (Table~A2), we compare the total wavenumber ($\sqrt{k^2_x+k^2_y}$) with the cutoff wavenumber ($k_c$). However, for coarse-graining, we have to compare $k_x$ and $k_y$ with $k_c$ (Eq.~\eqref{eq:filtered les 2d spectral velocity}). This subtle difference results in the Leonard stress not being zero here. Note that our conclusion about the failure of discovering a robust closure model with the sharp-spectral filter (at least with the current library and algorithm) is not affected by these choices. We have found that we still cannot discover a closure when we repeat the equation-discovery process on the total stress or the Reynolds stress calculated with filtering that also uses Eq.~\eqref{eq:filtered les 2d spectral velocity}.

\begin{figure}[t]
  \includegraphics[width=\textwidth]{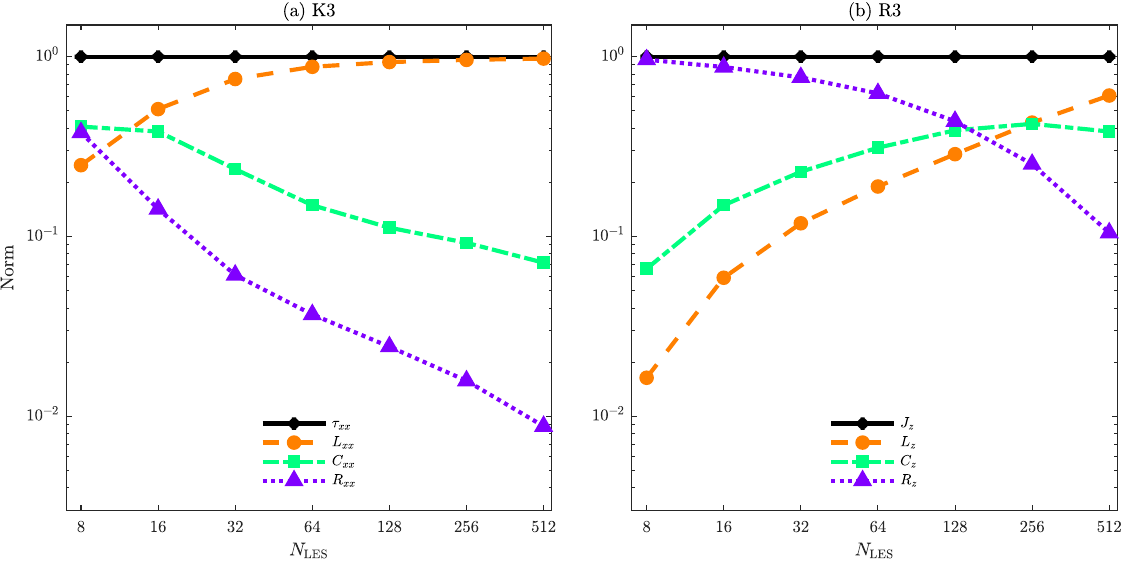}
    \centering
  \caption{The $L_2$-norm of the SGS components versus $N_{\text{LES}}$. (a) $\tau_{xx}$ from Case K3. (b) $J_z$ from Case R3. The contribution of SGS components is dependent on filter size: as $N_{\text{LES}}$ decreases, i.e., $\Delta$ increases, the relative importance of Reynolds stress (Leonard stress) increases (decreases). Norm of all the SGS components are normalized by the respective SGS flux's norm. A Gaussian filter is used, but the same behavior is observed for the box and Gauassian+box filters.}
  \label{fig: norm sgd components}
\end{figure}

As for the dependency on filter size, as $\Delta$ increases, the relative importance of Reynolds stress increases: See \cref{fig: norm sgd components} for examples from Cases K3 ($\tau_{xx}$) and R3 ($J_z$). Finally, note that the relative importance of these three components might depend on the flow itself. For example, in 3~km-resolution regional simulations of the tropics, \citeA{sun2023quantifying} found that the vertical (horizontal) flux of the SGS zonal momentum is dominated by the Reynolds (Leonard) stress, which was attributed to the substantial differences of the filtered vertical wind and the filtered zonal or meridional winds.

The above analyses further explain the strong dependency of the diagnosed ``true'' SGS flux and the discovered closures on the filter type and size. These analyses also show that depending on the filter type/size, the Reynolds stress may not be the only component of the SGS flux that needs to be parameterized. In fact, the Leonard and cross stresses might be even larger and have to be included in the calculation of the total SGS flux and in the closure. \KJ{Thus, there is no unique way to quantify the SGS fluxes, and} these sensitivities have major implications for the ``true'' SGS flux that is fed into the RVM or any equation-discovery algorithm (and more broadly, any ML algorithm). \KJ{Needless to say, these sensitivities have major impacts on the choices to make for developing data-driven closures in real-world applications (see item (f) in the next section).}  

\section{Summary and Discussion} \label{sec: summary and discussion}

In this work, we have used relevance vector machine (RVM) to discover subgrid-scale (SGS) closures from filtered direct numerical simulation (DNS) data for both the SGS momentum flux tensor (in 2D forced homogeneous isotropic turbulence, 2D-FHIT, and Rayleigh-B\'enard convection, RBC) and the SGS heat flux vector (in RBC). The expansive library includes derivatives of velocity (and temperature) up to 8th order (calculated using spectral methods) and their quadratic combinations. We have conducted extensive robustness analysis of the discovered closures across a variety of flow configurations (changing $Re, Ra, Pr,$ and the forcing wavenumber), filter types (Gaussian, box, Gaussian + box, and sharp-spectral cutoff), and filter sizes. 

Based on these analyses, except for when the sharp-spectral filter is used (see below), we have robustly discovered the {\it same closure} for the SGS stress in 2D-FHIT and RBC. We have further shown that this closure model is in fact the NGM, which can be derived analytically from the first term of the Taylor-series expansion of the convolution integral. The discovered SGS heat flux in RBC is also consistent with the truncated Taylor-series expansion. We have demonstrated a few important points about these discovered closures:
\begin{enumerate}
    \item They all have high CC (often $>0.9-0.95$) with the true SGS terms obtained from filtered DNS data, i.e., excellent performance based on this commonly used \apriori test metric. The same closure is discovered regardless of the system because the expansion's first term dominates the MSE loss function of RVM.
    \item Despite this high CC, all \aposteriori (online) tests result in unstable LES. This is consistent with the past findings about the NGM in the LES community (mainly for 3D turbulence) and in the climate community (for geophysical turbulence). Here, we argue that the inability of NGM to capture any inter-scale kinetic energy transfer in 2D-FHIT (or any 2D flow filtered in both directions) is likely the reason for the instability. For RBC, where filtering is done only in one direction, deeper investigations into the spectra of the SGS fluxes and inter-scale enstrophy and potential energy transfer, pointed to another likely reason for the instability: poor representation of the backscatter of SGS potential energy. This suggests that the poor {\it a posteriori} (online) performance of NGM might have different reasons in different flows.
    \item The exact form of the discovered closure depends on the filter type and the filter size, $\Delta$. For filters with compact support (i.e., all filters used here except for sharp-spectral cutoff), the structure of the closures is the same, but the coefficients are different (still, consistent with the Taylor-series expansion, as shown in the appendices). For the sharp-spectral cutoff filter, the equation discovery fails, again, consistent with the fact that the Taylor-series expansion cannot be conducted, a known issue in the LES literature \cite{sagaut2006large}. \KJnew{Finally, we would like to emphasize that what is reported here, e.g., based on low CC, is the ``failure'' of the ``equation discovery process'', whose goal was to minimize the MSE loss to match the stress (i.e., structural modeling). As discussed earlier in the paper as well as further below, functional closures with low CC (such as \citeA{smagorinsky1963general}) can produce stable and relatively reasonable LES, thus low CC does not necessarily mean a failed closure. That said, we could not discover any parsimonious closure with the sharp-spectral filtered data that was consistent across several cases and resolutions, and thus, we do not perform any \aposteriori LES with these low-CC closures.} Note that as mentioned before, with a ``smoothed'' sharp-spectral filter, \citeA{ross2023benchmarking} successfully performed equation-discovery.
\end{enumerate}
As a side note, while the terms of the discovered closures might look like truncation error of finite difference/volume discretization, in our work, all calculations (DNS solver, SGS terms, library) are done using spectral methods. This further shows, along with the Taylor-series expansion results, that the discovered closures are indeed representing the physics of the SGS terms, rather than any numerical error.

As an additional piece of analysis, we also present the decomposition of the SGS terms to the Leonard, cross, and Reynolds terms. We show that the Leonard and then cross terms often dominate the total SGS term, though the relative amplitude of these terms decreases as the filter size increases. However, this analysis shows that only computing the Reynolds momentum stress or heat flux can lead to discovering an inaccurate closure (and in general, in data-driven SGS modeling, in too-small SGS fluxes). That said, the relative importance of these 3 terms depends on the filter type and size, and likely, on the flow's spatial spectrum \cite{sun2023quantifying}. 

The analyses presented in this paper are aimed at highlighting the promises and challenges of the equation-discovery approach to SGS modeling. On one hand, it is promising that this approach robustly discovers closures that could be closely connected with those mathematically derived, and could be easily interpreted and analyzed in terms of turbulence physics. On the other hand,
\begin{itemize}
    \item[a)] The commonly used MSE loss function, or similar loss functions, will be always dominated by the leading term(s) of the Taylor-series expansion. Thus, sparsity-promoting equation-discovery techniques, at least with the common derivative/polynomial-based libraries, will always find the NGM (if all the relevant terms exists in the library). Note that this is true for the closure of any SGS process, as the Taylor-series expansion applies to any compact filter. {\it Thus, this point and many of the main points of this paper are relevant beyond just SGS modeling for turbulence, but also SGS modeling of other nonlinear, multi-scale processes in the Earth system}. 
    \item[b)] Given that our diagnoses show shortcomings of the NGM with functional modeling metrics (e.g., inter-scale energy transfer), one idea is to include such physics constraints in the loss function. For example, \citeA{guan2023learning} demonstrated that a loss function that combines structural and functional modeling constraints can enhance the \apriori and \aposteriori performance of the data-driven closure model in the small-data regime. More functional-modeling physics constraints (as domain averaged or wavenumber-dependent quantities) can be included in the loss function, which can potentially close the gap between {\it a priori} (offline) and {\it a posteriori} (online) performance. While the loss function of some techniques such as RVM may not be flexible to change beyond MSE, other methods such as \KJ{genetic programming, gene expression programming} or symbolic regression provide such flexibility \cite{ross2023benchmarking, cranmer2023interpretable}. Also, equation-discovery using neural network-based algorithms has gained popularity recently, as for example, their loss functions can be very flexible given the use of backpropagation for training \cite{chen2021physics}. That said, ``spectral bias'' \cite{xu2019frequency}, the fundamental challenge of neural networks in learning high wavenumbers, can become an issue when equation-discovery is the goal; see \citeA{mojgani2023interpretable} for an example and a solution in a quasi-geostrophic turbulence testcase.
    \item[c)] The fault may not entirely (or at all) lie with the MSE loss function. \citeA{guan2022stable} showed that a deep CNN with basically the same MSE loss function as the one used here (which only accounts for structural modeling) can learn a closure for 2D turbulence that has CC$>0.95$ and leads to stable and accurate LES (and accurate inter-scale transfers; see \citeA{guan2023learning}). But a major difference between the CNN and RVM is that the former does not use a pre-defined set of basis functions, but rather, {\it learns} them. Recent work by \citeA{subel2022explaining} has shown that the CNN of \citeA{guan2022stable} learned a set of low-pass, high-pass, and band-pass Gabor filters. As another major difference, the CNN's sparsity is not user-defined, but rather, comes from over-parameterization. 
    \item[d)] Related to (c), the discovered closures can depend on the choice of the library. This issue can be addressed by trying more expansive libraries (though this can lead to non-robust discoveries) or as mentioned earlier, by using methods such as \KJ{genetic programming} or \KJ{gene expression programming}, which allow the library to evolve (see \citeA{schmidt2009distilling,udrescu2020ai,ross2023benchmarking}). Libraries inspired by the CNNs' basis functions or distilled from other deep neural networks could be explored as well \cite{subel2022explaining, cranmer2020discovering}. Furthermore, there are studies, e.g., based on the Mori-Zwanzig formalism, suggesting that memory has to be included in closures 
    \cite{wouters2013multi,parish2017non}. Hence, basis functions that include temporal information (as already used in \citeA{ross2023benchmarking}) should be further explored in future work.
    \item[e)] Choosing the hyper-parameter(s) that determine the level of sparsity might require more thought too. While the L-curve criterion has shown success in many problems, the metrics for which the curve is calculated for should be further investigated. The common \apriori metrics such as CC of SGS fluxes are completely incapable of identifying shortcomings from a functional modeling perspective, such as lack of inter-scale energy transfer or poor representation of backscattering, which can be diagnosed using additional metrics. Note that a high CC of SGS fluxes has been found as the {\it necessary} but not {\it sufficient} condition for a successful closure \cite{meneveau1994statistics}.
    \item[f)] Aside from all of the above issues related to the discovery algorithm, what needs to be discovered (the ``truth'') should be further examined. The discovered closures can depend on the filter type/size and the methodology (e.g., calculating Reynolds stress or the full SGS stress), because what is diagnosed as the ``true'' SGS flux from DNS has such dependencies. This has important implications for any data-driven SGS modeling approach, including those using deep neural networks or any other statistical learning method \cite{fatkullin2004computational,zanna2021deep,grooms2021diffusion,guan2022stable,sun2023quantifying}. \KJ{In fact, this non-uniqueness and uncertainty of the true SGS term is a major shortcoming of the supervised/offline learning approach to data-driven closure modeling in real-world applications, and is one of the main motivations to pursue online or at least offline-online learning approaches \cite{schneider2021learning,schneider2023harnessing,pahlavan2024explainable}.} 
\end{itemize} 

We point out that there are other approaches to equation-discovery of SGS closures that are more directly focused on functional modeling. One is based on learning a closure from the differences between the evolved {\it states} of a high-resolution and a low-resolution simulation \cite{lang2016systematic, mojgani2022discovery, mojgani2023interpretable}. The other is to learn from the differences between the evolved long-term {\it statistics} of such simulations \cite{schneider2020imposing, schneider2021learning, schneider2022ensemble}. These approaches would partially or entirely resolve the issues (a), (b), and (f) mentioned above, although challenges (d) and (e) would remain. Furthermore, the {\it a priori} performance of such closures and challenges in interpretability arising from numerical errors accumulated during evolutions are left to be further investigated. 

In summary, equation-discovery is a promising approach to developing interpretable, practical, stable, and accurate SGS closures for various complex processes. However, further work, particularly on physics-guided loss functions (that for example, contain both structural and functional modeling components), physics- and data-guided libraries, and better metrics are needed.


\appendix

\section{Filtering Procedure} \label{appendix: filtering procedure}

\begin{figure}[t]
  \includegraphics[width=\textwidth]{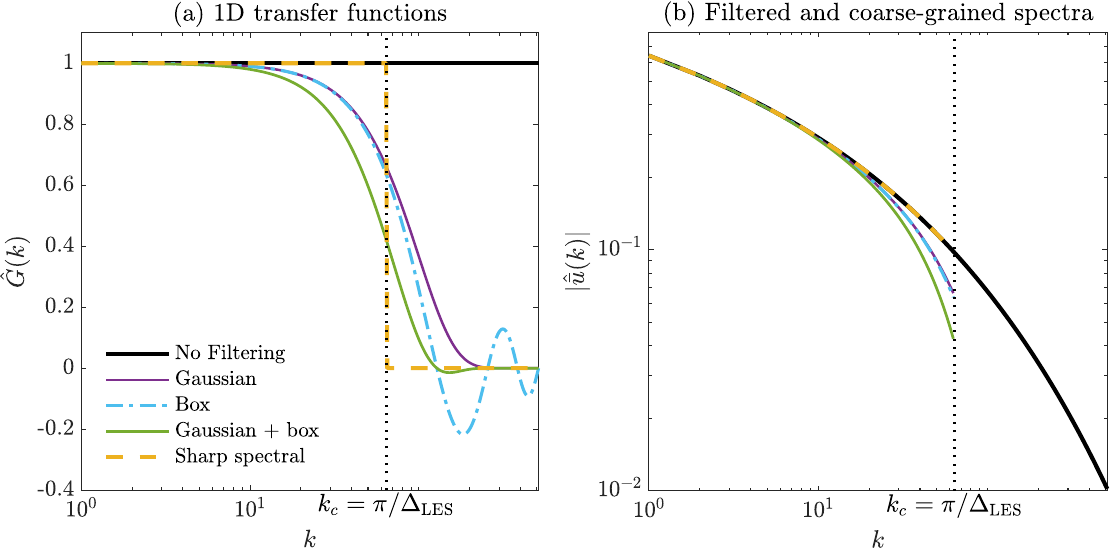}
    \centering
  \caption{(a) 1D transfer functions of filters listed in \cref{table:1d filters}. (b) The spectrum of a 1D filtered and coarse-grained variable. All filters are implemented in the spectral space, i.e., by applying their transfer function on Fourier-transformed variables. Here, \(k_c\) is the cutoff wavenumber for the sharp-spectral filter. Coarse-graining is performed in the spectral space using the same cutoff, \(k_c = \pi/\Delta_{\text{LES}}\).}
  \label{fig: 1D transfer function}
\end{figure}

In this work, we explore the most commonly used filters in LES and climate modeling: the Gaussian filter, the box filter, the Gaussian + box filter, and the sharp-spectral filter \cite{sagaut2006large,grooms2021diffusion}. The box filter (also known as the top-hat filter) is simply the average of a variable over a box of dimension $\Delta$; for instance, in 1D space, $\filtered{u}\left(x,t\right)$ 
is the average of $u\left(x_\circ,t\right)$ over $x-\Delta/2<x_\circ<x+\Delta/2$. 
The Gaussian filter's kernel is $G\left(r\right) = \frac{1}{\sigma\sqrt{2\pi}}\exp\left(-{\frac{1}{2}}\left(\frac{r-\mu}{\sigma}\right)^2\right)$, with zero mean, $\mu = 0$, and variance, $ \sigma^2 = \Delta^2/12$. These values are chosen to match the second moments (Eq. \eqref{eq:1D filter moments}) of the Gaussian and box filters following \citeA{leonard1975energy}. The kernel for the Gaussian + box filter is the convolution of the Gaussian and box filter kernels, which is equivalent to using a Gaussian filter followed by a box filter. The sharp-spectral cutoff filter simply removes the wavenumbers beyond a cutoff, $k_c$. The 1D filters used in this work are listed in \cref{table:1d filters}, and the 2D filters are listed in \cref{table:2d filters}. \KJ{\Cref{fig: 1D transfer function} illustrates the transfer functions of the 1D filters and their application to a variable.} Note that all of these 4 filters commute with the spatial and temporal derivative operators on uniform grids \cite{pope2000turbulent,sagaut2006large}. 

\begin{table}
\caption {List of \gls{1d} filters and their kernel and transfer functions. All filters are implemented in the spectral space, i.e., by applying their transfer function on Fourier-transformed variables. Here, $r$ and $k$ are coordinates in the physical space and spectral space, respectively. $\odot$ is the Hadamard product, $\hat{\left(.\right)}$ is the Fourier transform\KJ{, and $\Delta= 2\Delta_{\text{LES}}$, where $\Delta_{\text{LES}} = {L}/{N_{\text{LES}}}$ is the LES grid spacing.}}
\centering
\begin{tabular}{ |c|c|c| } 
\hline
Filter                             & Kernel function
& Transfer function \\
\hline \hline
General                            &  $G \left(r\right) $
&  $\hat{G}\left(k\right) =  \int_{-\infty}^{\infty} e^{i 2\pi kr} G\left(r\right) dr$                  \\
\hline
Gaussian $\left(G_{\text{G}}\right)$        &     
  $\left(\frac{6}{\pi \Delta^2}\right)^{\frac{1}{2}} \exp\left({-\frac{6r^2}{\Delta^2}}\right)$
& $\exp\left({-\frac{k^2\Delta^2}{24}}\right)$                          \\
\hline
Box $\left(G_{\text{B}}\right)$             &$
\begin{cases}
    \frac{1}{\Delta},& \text{if } r\leq \frac{\Delta}{2} \\
    0,              & \text{otherwise}
\end{cases}$
& $\frac{\sin(\frac{1}{2}k\Delta)}{\frac{1}{2}k\Delta}$ \\
\hline
Gaussian + box                    & $G_{\text{G}}\left(r\right)*G_{\text{B}}\left(r\right)$
& $\hat{G}_{\text{G}}\left(k\right) \odot \hat{G}_{\text{B}}\left(k\right)$ \\
$\left(G_{\text{GB}}\right)$ & & \\
\hline
Sharp-spectral cutoff
& \multirow{3}{*}{$\frac{\sin\left(\frac{\pi r}{\KJ{\Delta_{\text{LES}}}}\right)}{\pi r}$}
& \multirow{3}{*}{
$\begin{cases}
    1,& \text{if } \big(k_c - \left|k\right| \geq 0\big), \KJ{k_c = \pi/\Delta_{\text{LES}}}          \\
    0,              & \text{otherwise}
\end{cases}$} \\
$\left(G_{\text{S}}\right)$ & & \\
 & & \\
\hline
\end{tabular}
\label{table:1d filters}
\end{table}

\begin{table}
\caption {List of \gls{2d} filters and their kernel and transfer functions. All filters are implemented in the spectral space, i.e., by applying their transfer function on Fourier-transformed variables. Here, $\bm{r}$ and $\bm{k}$ are coordinates in the physical space and spectral space, respectively, with $ \bm{r} = \left(r_x,r_y\right)$, $|\bm{r}|^2 = r_x^2 + r_y^2$, $\bm{k} = \left(k_x,k_y\right)$, and $|\bm{k}|^2 = k_x^2 + k_y^2$. $\odot$ is the Hadamard product and $\hat{\left(.\right)}$ is the Fourier transform\KJ{, and $\Delta= 2\Delta_{\text{LES}}$, where $\Delta_{\text{LES}} = {L}/{N_{\text{LES}}}$ is the LES grid spacing.} }
\centering
\begin{tabular}{ |c|c|c| } 
\hline
Filter                             & Kernel function
& Transfer function \\
\hline \hline
General                            &  $G \left(\bm{r}\right) $
&  $\hat{G}\left(\bm{k}\right) =  \int_{-\infty}^{\infty} \int_{-\infty}^{\infty} e^{i 2\pi \left(k_xr_x+k_yr_y\right)} G\left(\bm{r}\right) d\bm{r}$                  \\
\hline
Gaussian $\left(G_{\text{G}}\right)$        &     
   $\frac{6}{\pi \Delta^2} \exp\left({-\frac{6\left|\bm{r}\right|^2}{\Delta^2}}\right)$
& $\exp\left({-\frac{\left|\bm{k}\right|^2\Delta^2}{24}}\right)$                          \\
\hline
Box $\left(G_{\text{B}}\right)$             &$
\begin{cases}
    \frac{1}{\Delta^2},& \text{if } \left(r_x,r_y\right)\leq \frac{\Delta}{2} \\
    0,              & \text{otherwise}
\end{cases}$
& $\frac{\sin(\frac{1}{2}k_x\Delta) \sin\left(\frac{1}{2}k_y\Delta\right)}{\left(\frac{1}{2}k_x\Delta\right) \left(\frac{1}{2}k_y\Delta\right)}$ \\
\hline
Gaussian + box                    & $G_{\text{G}}\left(\bm{r}\right)*G_{\text{B}}\left(\bm{r}\right)$
& $\hat{G}_{\text{G}}\left(\bm{k}\right) \odot \hat{G}_{\text{B}}\left(\bm{k}\right)$ \\
$\left(G_{\text{GB}}\right)$ & & \\
\hline
Sharp-spectral
& \multirow{3}{*}{$\frac{\sin\left(\frac{\pi\left|\bm{r}\right|}{\KJ{\Delta_{\text{LES}}}}\right)}{\pi\left|\bm{r}\right|}$}
& \multirow{3}{*}{
$\begin{cases}
    1,& \text{if } \big(k_c - \left|\bm{k}\right| \geq 0\big), \KJ{k_c = \pi/\Delta_{\text{LES}}}          \\
    0,              & \text{otherwise}
\end{cases}$} \\
$\left(G_{\text{S}}\right)$ & & \\
 & & \\
\hline
\end{tabular}
\label{table:2d filters}
\end{table}

\section{The 2D-FHIT Numerical Solver} \label{appendix: 2d-fhit numerical solver}
The numerical solver is the same as the one used in \citeA{guan2022stable}. Briefly, we solve Eqs.~\eqref{eq:2d-fhit uv continuity}-\eqref{eq:2d-fhit uv momentum} in the vorticity-streamfunction, $\omega-\psi$, formulation, where
\begin{eqnarray}
 \omega = \left(\nabla \times \bm{u} \right) \cdot \hat{\bm{z}}.
\end{eqnarray}
With this formulation, the governing equations are
\begin{eqnarray}
\nabla^2\psi & = & - \omega, \label{eq:2d-fhit psi omega}\\
\frac{\partial \omega}{\partial t} + \mathcal{N}\left(\omega,\psi\right) & = &  \frac{1}{Re}\nabla^2\omega -\chi \omega - f, \label{eq:2d-fhit psi omega momentum}
\end{eqnarray}
where $\mathcal{N}(\omega,\psi)$ is 
\begin{eqnarray}{\label{eq:2d-fhit advection}}
    \mathcal{N}(\omega,\psi) = \frac{\partial \psi}{\partial y}\frac{\partial \omega}{\partial x} - \frac{\partial \psi}{\partial x}\frac{\partial \omega}{\partial y}.
\end{eqnarray}
$f$ is a deterministic forcing \cite{chandler2013invariant, kochkov2021machine}:
\begin{eqnarray}{\label{eq:2d-fhit forcing}}
    f = f_{k_x}\cos(f_{k_x}x) + f_{k_y}\cos(f_{k_y}y),
\end{eqnarray}
where $f_{k_{x}}$ and $f_{k_{y}}$ are the forcing wavenumbers and $\chi=0.1$ represents the Rayleigh drag coefficient. Comparing Eq. \eqref{eq:2d-fhit uv momentum} with Eq. \eqref{eq:2d-fhit psi omega momentum}, it is evident that $\nabla \times \bm{\mathcal{R}} = -\chi \omega$ and $\nabla \times \bm{\mathcal{F}} = -f$.

In DNS, Eqs.~\eqref{eq:2d-fhit psi omega}-\eqref{eq:2d-fhit psi omega momentum} are solved in a doubly periodic domain using a Fourier-Fourier pseudo-spectral solver with second-order Adams-Bashforth and Crank Nicholson time-integration schemes for the advection and viscous terms, respectively (time step $\Delta t_{\text{DNS}}$).  
For LES, we use the same solver with lower spatio-temporal resolution: We use $N_{\rm{LES}}$ that is 8 to 64 times smaller than $N_{\text{DNS}}$, and $\Delta t_{\text{LES}}=10\Delta t_{\text{DNS}}$.

\section{The RBC Numerical Solver} \label{appendix: rbc numerical solver}
We solve Eqs.~\eqref{eq:rbc uv continuity}-\eqref{eq:rbc uv energy} using a Fourier-Chebyshev pseudo-spectral solver \cite{khodkar2019reduced, khodkar2018data}. Briefly, using the $\omega-\psi$ formulation, the dimensionless governing equations become
\begin{eqnarray}
    \nabla^2 \psi & = & -\omega \label{eq:rbc psi omega} \\
    \frac{\partial \omega}{\partial t} + \mathcal{N}\left(\omega,\psi\right) & = & Pr \nabla^2\omega + Pr\,Ra\,\theta \hat{\bm{z}} \label{eq:rbc psi omega momentum}, \\
    \frac{\partial \theta}{\partial t} + \mathcal{M}\left(\theta,\psi\right) + \frac{\partial \psi}{\partial x} & = &\nabla^2 \theta, \label{eq:rbc psi omega energy}
\end{eqnarray}
where $\mathcal{N}(\omega,\psi)$ and $M\left(\theta, \psi\right)$ are
\begin{eqnarray}{\label{eq:rbc advection}}
    \mathcal{N}(\omega,\psi) = \frac{\partial \psi}{\partial z}\frac{\partial \omega}{\partial x} - \frac{\partial \psi}{\partial x}\frac{\partial \omega}{\partial z},
    \hspace{1em}
    \mathcal{M}(\theta,\psi) = \frac{\partial \psi}{\partial z}\frac{\partial \theta}{\partial x} - \frac{\partial \psi}{\partial x}\frac{\partial \theta}{\partial z}.
\end{eqnarray}

For DNS, we solve Eqs. \eqref{eq:rbc psi omega}-\eqref{eq:rbc psi omega energy} in domain $\left(6 \pi, 1\right) $. Periodic boundary conditions are imposed in the horizontal direction and no-slip and fixed temperature boundary conditions are imposed on the horizontal walls. Second-order Adams-Bashforth and Crank Nicholson time integration schemes are used for the advection and viscous terms, respectively. \cref{table:rbc cases} presents the $N_{\text{DNS}}$ and $N_{\text{LES}}$ for each case. For LES, we use the same solver but with lower resolution in the horizontal direction.

\section{Closure Discovery for Sharp-Spectral Filter} \label{appendix: sharp-spectral}
\begin{figure}[!h]
  \centering
  \includegraphics[width=\textwidth]{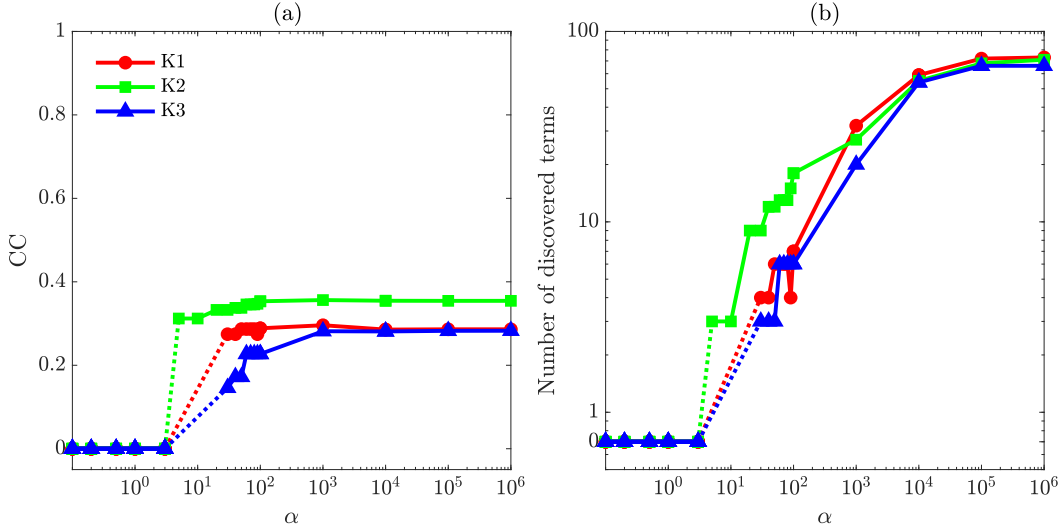}
  \caption{Representative examples of the failure of equation discovery when applied to the stress diagnosed using sharp-spectral filters. The panel shows the effects of increasing the sparsity-level hyper-parameter, $\alpha$, on the (a) CC and (b) number of terms in the discovered closure of $\tau_{yy}$ for 2D-FHIT. A sharp-spectral filter with $N_{\text{LES}}=128$ (for cases K1-K3) is used, but the same behavior is observed with any other $N_{\text{LES}}$ for the sharp-spectral filter. For this filter, equation discovery fails, leading to low CC and non-robust closures.
 }
  \label{fig: CC number of discovered terms sharp spectral}
\end{figure}

\KJ{The closures discovered by the sharp-spectral filter have low CC ($<0.4$, see \cref{fig: CC number of discovered terms sharp spectral}). Even increasing the complexity of the discovered closure by including more terms (\cref{fig: CC number of discovered terms sharp spectral}(b)) does not improve the CC. This stands in contrast to the high CC ($>0.95$) observed with the Gaussian, box, and Gaussian + box filters with only a few terms included. As a result, the equation discovery fails for the sharp-spectral filter.}

\section{Taylor-series Expansion of the SGS Flux for a 1D Field} \label{appendix: GM derivation 1d}
Let's focus on a spatially 1D field $a(x)$ (dependence on $t$ is not explicitly written for brevity). The filtering operation's convolution integral (Eq.~\eqref{eq:general filtering}) becomes
\begin{eqnarray} \label{eq:1dfilter}
    \filtered{a}(x) = G*a\ = \int_{-\infty}^{\infty} G\left(r_x\right) a\left(x-r_x\right) dr_x,
\end{eqnarray}
The Taylor-series expansion of $a(x-r_x)$ around $a(x)$ gives
\begin{eqnarray} \label{eq:1D taylor}
    a\left(x-r_x\right) = a\left(x\right) - \frac{1}{1!}\frac{\partial a\left(x\right)}{\partial x}r_{x} + 
    \frac{1}{2!}\frac{\partial^2 a\left(x\right)}{\partial x^{2}}r_{x}^{2} +
    \dots
\end{eqnarray}
Substituting this into Eq.~\eqref{eq:1dfilter} and using $a = a(x)$, $\filtered{a} = \filtered{a}(x)$ for brevity yields
\begin{eqnarray}\label{eq:1D filter taylor}
\filtered{a} & = & 
\int_{-\infty}^{\infty} G\left(r_x\right)a dr_{x} - 
\int_{-\infty}^{\infty} G\left(r_x\right)\frac{\partial a}{\partial x}r_{x} dr_{x} +
\frac{1}{2!}\int_{-\infty}^{\infty} G\left(r_x\right)\frac{\partial^2 a}{\partial x^{2}}r_{x}^{2} dr_{x} + \dots  \\
& = & 
a\int_{-\infty}^{\infty} G\left(r_x\right) dr_{x} - 
\frac{\partial a}{\partial x}\int_{-\infty}^{\infty} G\left(r_x\right)r_{x} dr_{x} +
\frac{1}{2!}\frac{\partial^2 a}{\partial x^{2}}\int_{-\infty}^{\infty} G\left(r_x\right) r_{x}^{2} dr_{x} + \dots \label{eq:afiltered}
\end{eqnarray}
The second line follows the first line considering that $a$ and its derivatives do not depend on the variable of integration, $r_x$. In Eq.~\eqref{eq:afiltered}, $\filtered{a}$ depends on $a$ and its derivatives, with coefficients that only depend on the filter type and size through moments of the kernel, $G$. For example, for a Gaussian filter (Table~\ref{table:1d filters}) 
\begin{eqnarray}\label{eq:1D filter moments}
    \int_{-\infty}^{\infty} G\left(r_x\right) dr_{x} = 1, 
    \int_{-\infty}^{\infty} G\left(r_x\right) r_{x} dr_{x} = 0, 
    \int_{-\infty}^{\infty} G\left(r_x\right) r_{x}^{2} dr_{x} = \frac{\Delta^2}{12}.
\end{eqnarray}
Note that all the odd moments are $0$, resulting in $\mathcal{O}\left(\Delta^4\right)$ as the order of the truncated terms once the moments in Eq.~\eqref{eq:1D filter moments} are substituted in Eq.~\eqref{eq:afiltered}: 
\begin{eqnarray}\label{eq:1D filtered u}
 \filtered{a} = a + 
\frac{1}{2!}\frac{\Delta^2}{12}\frac{\partial^2 a}{\partial x^{2}} + 
\mathcal{O}\left(\Delta^4\right).
\end{eqnarray}
To calculate a term like $\tau_{xx} = \filtered{u^2} - \filtered{u}^{2}$, we first use $a=u$ in Eq.~\eqref{eq:1D filtered u} and then square it to arrive at 
\begin{eqnarray}\label{eq:1D filtered u^2}
\filtered{u}^{2} = u^{2} + 
2u\left(\frac{1}{2!}\frac{\Delta^2}{12}\frac{\partial^2 u}{\partial x^2}\right) 
+ \mathcal{O}\left(\Delta^4\right).
\end{eqnarray}
Next, we use $a=u^2$ in Eq.~\eqref{eq:1D filtered u} to obtain
\begin{eqnarray}
\filtered{u^{2}} & =  & u^{2} + 
\frac{1}{2!}\frac{\Delta^2}{12}\frac{\partial^2 u^{2}}{\partial x^2} + 
\mathcal{O}\left(\Delta^4\right), \\
& = &
u^{2} + 
\frac{2}{2!}\frac{\Delta^2}{12}\left(\left(\frac{\partial u}{\partial x}\right)^{2}+
u\frac{\partial^2 u}{\partial x^2}\right) 
+ \mathcal{O}\left(\Delta^4\right). \label{eq:1D filtered^2 u}
\end{eqnarray}
Using Eq.~\eqref{eq:1D filtered u^2} and Eq.~\eqref{eq:1D filtered^2 u} we find
\begin{eqnarray}\label{eq:1D tau 1}
\tau_{xx} = \filtered{u^2} - \filtered{u}^{2} = 
\frac{\Delta^2}{12}\left(\frac{\partial u}{\partial x}\right)^{2} + \mathcal{O}\left(\Delta^4\right).
\end{eqnarray}
Note that this expression depends on $u$ rather than $\filtered{u}$, which is what we desire. Next, we use $a=\partial u/\partial x$ in Eq.~\eqref{eq:1D filtered u} to obtain
\begin{eqnarray}\label{eq:1D filtered ux}
\frac{\partial \filtered{u}}{\partial x} = 
\frac{\partial u}{\partial x} + 
\frac{1}{2!}\frac{\Delta^2}{12}\frac{\partial^3 u}{\partial x^3} + 
\mathcal{O}\left(\Delta^4\right).
\end{eqnarray}
Using this expression in Eq.~\eqref{eq:1D tau 1} yields an analytically derived closure for $\tau_{xx}$ with error $\mathcal{O}\left(\Delta^4\right)$ 
\begin{eqnarray}
\tau^{\rm{NGM}}_{xx} = \filtered{u^2} - \filtered{u}^{2} = 
\frac{\Delta^2}{12}\left(\frac{\partial \filtered{u}}{\partial x}\right)^{2}. \label{eq:GM21D}
\end{eqnarray}
This is the NGM \cite{leonard1975energy, sagaut2006large}. Four issues should be emphasized here
\begin{itemize}
\item [i.] \KJ{This procedure can be followed for any filter type. However, the Taylor series does not exist for filters such as sharp-spectral (cutoff), whose kernel's second-order moment is indefinite. Thus, for such filters, NGM does not exist~\cite{meneveau2000scale,sagaut2006large}}. 
\item  [ii.] The same procedure can be followed to derive NGM for higher dimensions, e.g., $\tau^{\rm{NGM}}_{xx}$, $\tau^{\rm{NGM}}_{xy}$, and $\tau^{\rm{NGM}}_{yy}$ in 2D; see \citeA{sagaut2006large}.
\item [iii.] The coefficients in NGM depend on the filter's kernel and its moments (Eq.~\eqref{eq:1D filter moments}). For Gaussian and top-hat, the parameters of the kernels are chosen to match their first moment, leading to $\Delta^2/12$ coefficient for both. However, the coefficients differ for higher-order terms \cite{sagaut2006large}.
\item [iv.] The procedure presented above is not specific to turbulence or even dynamical systems. The procedure and its outcome are valid for the filtered quadratic nonlinearity of any two variables, even random variables. 
\end{itemize}

\section{Subgrid-scale Energy and Enstrophy Transfers} \label{appendix: sgs energy transfer}
The filtered rate of train tensor $\filtered{\textbf{S}}$ and the anisotropic part of the SGS stress tensor $\bm{\uptau}^{\text{r}}$ are
\begin{eqnarray} \label{eq: filtered S}
    \filtered{\textbf{S}} & = &
    \begin{bmatrix}
    \frac{\partial \filtered{u}}{\partial x} & 
    \frac{1}{2}\left(\frac{\partial \filtered{u}}{\partial y} + \frac{\partial \filtered{v}}{\partial x} \right) \\ 
    \frac{1}{2}\left(\frac{\partial \filtered{u}}{\partial y} + \frac{\partial \filtered{v}}{\partial x}\right) &
    \frac{\partial \filtered{v}}{\partial y}
    \end{bmatrix},\\
    \bm{\uptau}^{\text{r}} & = & \bm{\uptau} - \frac{1}{2}\text{tr}\left(\bm{\uptau}\right) \bf{I}, 
    \label{eq:taur}
\end{eqnarray}
where $\bf{I}$ is the identity matrix. In 2D with filtering in both directions, the anisotropic part of the SGS stress tensor from the NGM is
\begin{eqnarray}
    \bm{\uptau}^{\text{NGM-r}}& = & \bm{\uptau}^{\text{NGM}} - \frac{1}{2}\text{tr}\left(\bm{\uptau}^{\text{NGM}}\right) \bf{I}. \label{eq:tauGMr}\\
    \bm{\uptau}^{\text{NGM-r}}_{\text{2D}} & = & \frac{\Delta^2}{12}
    \begin{bmatrix}
    \frac{1}{2}\left(\left(\frac{\partial \filtered{u}}{\partial y} \right)^2 -
    \left(\frac{\partial \filtered{v}}{\partial x} \right)^2\right)&
    \frac{\partial \filtered{u}}{\partial x}\frac{\partial \filtered{v}}{\partial x} + 
    \frac{\partial \filtered{u}}{\partial y}\frac{\partial \filtered{v}}{\partial y} \\
    \frac{\partial \filtered{u}}{\partial x}\frac{\partial \filtered{v}}{\partial x} + 
    \frac{\partial \filtered{u}}{\partial y}\frac{\partial \filtered{v}}{\partial y} &
    -\frac{1}{2}\left(\left(\frac{\partial \filtered{u}}{\partial y} \right)^2 -
    \left(\frac{\partial \filtered{v}}{\partial x} \right)^2\right)
    \end{bmatrix}. \label{eq:tauGMr-2D}
\end{eqnarray}
Inserting this and $\filtered{\textbf{S}}$ (Eq.~\eqref{eq: filtered S}) into Eq.~\eqref{eq: sgs P} shows zero point-wise inter-scale (kinetic) energy transfer in NGM: $P^{\text{NGM}}_{\tau}(x,y,t) = 0$.

In buoyancy-driven turbulence such as RBC, the total inter-scale energy transfer rate $P_E$ is the sum of the rate of transfer of kinetic energy ($P_{\tau}$) due to SGS momentum fluxes and potential energy ($P_{J}$) due to SGS heat fluxes \cite{eidson1985numerical,peng2002subgrid}:
\begin{eqnarray}
    P_{E} & = & P_{\tau} + P_{J} \nonumber \\
    & = & -\tau^r_{ij}\filtered{S}_{ij} - 
    Ra\, Pr\, J_z.
\end{eqnarray}
Given the 1D filtering used in RBC, $\bm{\uptau}^{\text{NGM-r}}_{\text{1D}}$ becomes
\begin{eqnarray}
    \bm{\uptau}^{\text{NGM-r}}_{\text{1D}} = 
    \frac{\Delta^2}{12}
    \begin{bmatrix}
    \frac{1}{2}\left(\left( \frac{\partial \filtered{u}}{\partial x}\right)^2 - \left(\frac{\partial \filtered{w}}{\partial x} \right)^2 \right) &
    \frac{\partial \filtered{u}}{\partial x}\frac{\partial \filtered{w}}{\partial x} \\
    \frac{\partial \filtered{u}}{\partial x}\frac{\partial \filtered{w}}{\partial x} &
    -\frac{1}{2}\left(\left( \frac{\partial \filtered{u}}{\partial x}\right)^2 - \left(\frac{\partial \filtered{w}}{\partial x} \right)^2 \right) 
    \end{bmatrix},
\end{eqnarray}
and $P_{\tau}^{\text{NGM}}$ is not strictly zero: 
The resulting production of \gls{sgs} energy transfer for NGM is \begin{eqnarray}
    P_{\tau}^{\text{NGM}}&= &-\frac{\Delta^2}{12} \left(\frac{\partial^3 \filtered{u}}{\partial x^3}+\frac{\partial \filtered{u}}{\partial x}\frac{\partial \filtered{u}}{\partial z}\frac{\partial \filtered{w}}{\partial x} \right). \\
    P_J^{\text{NGM}} &= &-Ra \, Pr \,\frac{\Delta^2}{12} \frac{\partial \filtered{w}}{\partial x}\frac{\partial \filtered{\theta}}{\partial x}
\end{eqnarray}
 
Similarly, one can define the inter-scale enstrophy transfer for 2D-FHIT and RBC as \cite{chen2003physical}
\begin{eqnarray}
    P_Z = -\nabla{\filtered{\omega}} \cdot \left(\filtered{\bf{u} {\omega}}-\filtered{\bf{u}} \, \filtered{\omega}\right).
\end{eqnarray}

\section*{Open Research}
\KJ{The code for 2D-FHIT solver ``py2d'' is available at \citeA{py2d}. The code and data used for analysis in this work} can be found at \citeA{eqsdiscovery} and  \citeA{eqsdiscovery_data}, respectively. 

\acknowledgments
We thank \KJ{Laure Zanna for extensive and insightful discussions throughout this study. \KJ{We express our gratitude to Malte Jansen and two anonymous reviewers, as well as to the editor (Tapio Schneider), for their constructive feedback and suggestions.}  We are also grateful to} Ian Grooms, Sina Khani, Tony Leonard,  Charles Meneveau, Alistar Adcroft, and Pavel Perezhogin for their thoughtful comments and suggestions. We thank Moein Darman, Hamid Pahlavan, and Qiang Sun for their helpful comments on the manuscript. This work was supported by an award from the ONR Young Investigator Program (N00014-20-1-2722), a grant from the NSF CSSI program (OAC-2005123), and by the Schmidt Sciences, LLC. Computational resources were provided by NSF XSEDE (allocation ATM170020) and NCAR's CISL (allocations URIC0004 and URIC0009).

\bibliography{main}

\end{document}